 \documentclass[12pt,preprint]{aastex}

\shorttitle{A Simple Bipolar Jet Model for the Polarization
            of Core-Collapse Supernovae}
\shortauthors{David J. Jeffery}

\begin{document}


\title{A Simple Bipolar Jet Model for the Polarization
            of Core-Collapse Supernovae}


\author{David J. Jeffery}
\affil{Physics Department, University of Nevada, Las Vegas, Box 454002,
        4505 Maryland Parkway Las Vegas, Nevada 89154, U.S.A.} 
\email{jeffery@physics.unlv.edu}





\begin{abstract}
      We propose a bipolar jet model for supernova polarization.
Light from the main component of the supernova (which we
call the bulk supernova for short)
scatters off electrons in the jets and is polarized:  this polarized light 
is added to the direct emission from the bulk supernova and causes the
overall supernova emission to be polarized. 
The motivation for the bipolar jet model is to resolve the potential
paradox set by an analysis which suggests strong spherical
symmetry for Type~Ib supernovae and the expectation that Type~Ib
supernovae should be highly polarized in the continuum
up to perhaps $\sim 4\,$\%.
For the model we have derived simple, non-relativistic,
approximate analytic formulae for relative amount of scattered flux 
and polarization.
The polarization formula qualitatively reproduces two main
features observed for at least some supernovae:
(1)~the rise to a polarization maximum and then decline with
time;  (2)~the inverted P~Cygni polarization profiles of lines.
The model is strictly axisymmetric and predicts a constant
position angle of intrinsic polarization.
Since position angle variation of intrinsic polarization is observed, 
the model cannot account for all the polarization of all supernovae.
\end{abstract}


\keywords{supernovae: general---polarization---radiative transfer}


\section{Introduction}

\subsection{Supernova Polarization in General}

     Evidence has accumulated that net emission from core-collapse supernovae
is often, perhaps usually (or maybe even nearly always),
intrinsically polarized.
So far the observations show a range in intrinsic continuum polarizations
from $\sim 0\,$\% up to perhaps $4\,$\% with of order $1\,$\% being
typical
(e.g., \citealt{wang1996, wang2001}; \citealt{wheeler2000};
\citealt{leonard2000a, leonard2000b, leonardfilippenko2001, leonard2001}).
(The $\sim 4\,$\% polarization was actually obtained for
a Type Ic supernova [\citealt{wang2001}, but we assume it here as
a tentative general upper limit on core-collapse supernova continuum
polarization.)
P~Cygni line trough polarization features can exhibit higher
polarization than the continuum:  at the moment $4\,$\% for Ca~II line
trough of SN~1987A seems to be the record (\citealt{jeffery1991b}, Fig.~16).
The apparent polarization trends are that polarization increases with
time from explosion for perhaps up to of order 200 days or more and that it
increases with decreasing hydrogen/helium envelope mass of progenitor
at the time of explosion.
The sequence of decreasing hydrogen/helium envelope mass in
core-collapse supernova types
is II-P, 
II-L,    
II-n,    
IIb,     
Ib,      
and Ic
\citep{leonard2001}.  
The order of II-L, II-n, and IIb is 
uncertain.      
For an explanation of the supernova classification scheme see
reviews by, e.g.,
\citet{filippenko1997}, \citet{wheeler2000}, and \citet{cappellaro2001}.
(Type~Ia supernovae [thermonuclear supernovae] seem to be
usually much less polarized or unpolarized [\citealt{wang1996,
wheeler2000}] although some detections have been reported
[\citealt{wang1997, leonard2000b, howell2001}].
We will not discuss Type~Ia supernovae further in this paper.)

     The presence of intrinsic polarization shows some sort of asymmetry
exists in the supernova ejecta or the near-supernova environment
and that there is some sort polarizing scattering off electrons
in the ejecta or, perhaps, dust well away from the actual ejecta.
The continuum opacity of supernova atmospheres in the photospheric epoch
in the optical and infrared (IR) is in fact dominated by 
electron scattering opacity (e.g., \citealt{wagoner1981};
\citealt{harkness1991}), and thus the atmospheres themselves
would yield a net polarization if they were asymmetric somehow.
Ellipsoidal atmospheres models exploiting electron scattering
have been extensively explored
(e.g., \citealt{shapiro1982}; \citealt{mccall1984,mccall1985};
\citealt{jeffery1991a}; \citealt{hoeflich1991};
and \citealt{howell2001}).
Although, electron scattering has virtually no
wavelength dependence for photons with energy much less than
the electron rest mass energy
(e.g., \citealt{davisson1965}, p. 51--52), 
strong line polarization features caused by the
combination of electron scattering and line radiative transfer
can occur in these models through a mechanism investigated
by \citet{jeffery1989}.
Another asymmetric supernova model invokes clumps in the
ejecta as the cause of asymmetry (e.g., \citealt{chugai1992}).

     There is a model of supernova polarization that does not require
an asymmetric supernova.
Instead it invokes a relatively
remote scattering from a polarizing dust cloud \citep{wangwheeler1996}.
Because the cloud is remote the scattered spectrum is added
to the direct supernova spectrum of a later epoch:  the
scattered spectrum is time-delayed.
Dust scattering has a relatively weak polarization wavelength
dependence (e.g., \citealt{serkowski1973}), but the time-delay
effect can give rise to line polarization features due to the variation
of the line profiles over the time delay.
(In the bipolar jet model the line polarization features
arise through a significant redshift of the jet-scattered flux 
from its original observer frame wavelength.
We call this the redshift effect:  see \S\S~3.3--3.5 and~4).

     A considerable problem in analyzing intrinsic
supernova polarization is contamination by interstellar
polarization (ISP) caused by intervening dust along the line
of sight from the supernova.
That intrinsic polarization is present for almost all
core-collapse supernovae is proven by the time dependence of the polarization
(which would not happen for pure ISP) and the non-ISP wavelength dependence
of polarization in particular across lines.
(The ISP wavelength dependence is slowly varying and
follows the empirical Serkowski law:  e.g., \citealt{serkowski1973}.)
Unfortunately, ISP can completely distort an 
intrinsic polarization spectrum:  e.g., polarization maxima
can be turned into minima and vice versa.
The correction for ISP must be based on the observed supernova 
data itself and, alas, on usually somewhat uncertain assumptions.
Approaches to ISP correction are discussed by, e.g.,
\citet{jeffery1991b}, \citet{leonard2000a}, and \citet{wang2001}.

\subsection{A Supernova Polarization Paradox and the Bipolar Jet Model}

     Type~Ib supernovae (SNe~Ib) lack conspicuous hydrogen lines
although most or all may exhibit weak hydrogen Balmer
lines (e.g., \citealt{branch2002}, hereafter BBK).
These supernovae may be essentially helium cores of a few solar masses
that result from mass loss by massive stars ($\gtrsim 8\,M_{\odot}$
on the main sequence)
to a binary companion or in strong winds (e.g., \citealt{woosley1994}, 
p.~122ff). 
The hydrogen remaining may be a few tenths of a solar mass
or less which in the supernova ejecta is on the outside moving at
velocities $\gtrsim 11,000\,{\rm km/s}$ (BBK).
(Type~Ic supernovae are thought to be similar to SNe~Ib, but they
have lost even more envelope and show no conspicuous optical
helium lines although some helium remains in some cases at least
[\citealt{filippenko1995}; \citealt{clocchiatti1996}]).
The rise time to $V$~maximum for low-envelope-mass core-collapse supernovae 
like SNe~Ib is of order 20~days as
suggested by earliest observations (e.g., BBK)
and theoretically (e.g., \citealt{shigeyama1994, baron2001}).

    Recently, BBK have presented a parameterized LTE analysis of SN~Ib spectra.
Out of a small, but perhaps representative, sample of 6 SNe~Ib
BBK find that 5 show photospheric velocity curves (i.e., 
plots of the photospheric velocity evolution with time) 
that for the most part vary by 
$\lesssim 20\,$\% from a mean photospheric velocity curve
for up to about the first 80~days after explosion.
Since photospheric velocity evolution is in fact a function of 
many properties of a supernova (e.g., mass, density distribution,
composition, thermal state), the similar-evolution result
strongly suggests that 5 out of 6~supernovae in this sample are quite
alike:  tentatively the result suggests that most SNe~Ib could
be quite alike.
Moreover, the members of the subsample of 5 (hereafter the BBK subsample)
are probably fairly spherically symmetric
or else different viewing orientations 
would give different photospheric velocity evolutions even if the supernovae 
were identical.

     Recall now that
SNe~Ib according to the cited trend of
increasing polarization with decreasing envelope mass and some observations 
\citep{mccall1985, leonard2000b}
should be highly polarized and have intrinsic continuum polarizations of
order $1\,$\% or perhaps sometimes more.
To obtain polarization of $\gtrsim 1\,$\% from models
with ellipsoidal density distributions requires
for major-over-minor axis ratios $\gtrsim 1.2\,$\% 
in the iso-density contours (e.g, \citealt{hoeflich1991,hoeflich2001}).
Such axis ratios suggest $\gtrsim 20\,$\% variations in photospheric velocity
curves for identical supernovae seen at the same epoch at different 
viewing orientations.
It seems marginally unlikely that BBK subsample could exhibit the expected
polarizations if ellipsoidal asymmetry is assumed.

     Unfortunately, only SN~1983N out of the BBK subsample
has been measured for polarization.
Spectropolarimetry for SN~1983N was briefly reported on by \citet{mccall1985}.
(The SN~1983N spectropolarimetry data has never been published
and now may not be recoverable [\citet{mccall2001}].)
A polarization dip from $1.4\,$\% to $0.8\,$\% was found across
an Fe~II blended line flux emission near $4600\,$\AA\ near
$V$~maximum:  the position angle of polarization stayed constant
across the emission.
The change in polarization across a line profile is a clear
signature that intrinsic polarization was present, not just ISP
(see \S~1.1). 
The lack of variation in position angle suggests, in fact,
that nearly all the polarization was intrinsic and thus that
the intrinsic continuum polarization may have been as high as $1.4\,$\%.
(Any significant change in the relative amounts of 
intrinsic polarization and ISP would tend to cause a shift in position
angle, unless coincidentally the intrinsic and ISP position angles 
were aligned.)
Thus, at least one of the BBK subsample is 
significantly polarized and it is the possible they all are.

     We face the potential paradox of fairly spherically
symmetric supernovae yielding high polarization.
To resolve this paradox, we propose
a bipolar jet model for the polarization of core-collapse supernovae.
In the model, bipolar jets are thrown out of the supernova at the time 
of explosion.
If the jets are thrown out of the supernova core, it is plausible that
they are enriched in radioactive $^{56}$Ni.
The explosion is such that these jets are well collimated and
the bulk of the ejecta (which for short we will call the bulk supernova) 
is left fairly spherical:  a contrived
picture, but it is what we need for the resolution of the paradox.
Since the bulk supernova is fairly spherically
symmetric there is no viewing orientation variation in photospheric
velocity provided one is not looking nearly down a jet axis.
The jets stay sufficiently ionized (perhaps due to radioactive
decay of $^{56}$Ni and its daughter $^{56}$Co)
so that their electron optical depth is high enough 
that light emitted by the photosphere of the bulk supernova
and scattered by the jets into the line of sight is highly polarized.
The photosphere and jet are sufficiently remote that
the polarization position angle of the scattered light is nearly
constant and perpendicular to the plane defined by
the jet axis and the line of sight.
(See the discussion of the polarization properties of
electron scattering in \S~3.3.)
This means that the polarization of the scattered light
is not completely canceled and may amount to a polarization of
tens of percent.
The polarized scattered light from the jets is only a small
contribution to the net supernova flux, but as we will
show in \S~4 that contribution
can give the net flux a significant overall polarization:
e.g., $\gtrsim 0.5\,$\%.

     The existence of bipolar jets originating
in core-collapse events has been much investigated recently
and models have been calculated that show their effects
(e.g, \citealt{hoeflich2001, macfadyen2001, zhang2003}).
The jets have been inserted into these models as parameterized entities.
It is not clear if they arise naturally in core collapses.
Thus, we are free to invoke bipolar jets with the properties
we require without justifying them by the results of 
hydrodynamic calculations.

     In \S~2 of this paper, we present the structure of
our bipolar jet model.
Section~3 describes the radiative transfer in the model
and derives simple analytic formulae for relative amount of
jet-scattered flux (i.e., the relative scattered flux) and the
supernova polarization.
We assume non-relativistic radiative transfer as a simplifying hypothesis.
Because the jet is somewhat remote from the supernova, the 
scattered spectrum is added to the direct supernova spectrum of a 
later epoch: 
thus there is a time-delay effect that also occurred in the dust cloud model
of \citet{wangwheeler1996}.
The time-delay effect could be included in our model, and 
may sometimes be important even for non-relativistic jets.
We also allow for time evolution as the jets expand.
Our simplifications allow us to capture the main effects of 
non-relativistic bipolar jets including time evolution.
There are 5 independent basic parameters for the bipolar jet model:  
$\theta$ (or alternatively $\mu$), 
$\Delta\theta$, $\tau_{\rm rad}$, $\beta_{\rm in}$, and $b$.
The $\tau_{\rm rad}$ parameter can be reparameterized in terms
of physical parameters for the jet (see \S~2).
In a calculation in which the time-delay effect is included, a reduced time,
$t_{\rm red}$ (see \S~3.2),
replaces $\tau_{\rm rad}$ as one of the basic parameters.
In a calculation of the time evolution of polarization,
the parameter $\tau_{\rm rad,0}t_{0}^{2}$ replaces $\tau_{\rm rad}$
(see \S~2).
These parameters are all explained in \S\S~2, 3 and~4:  see especially
\S~3.5.

     Section~4 reports some demonstrations calculations obtained using our
analytic formulae.
Conclusions and final discussion are given in \S~5.

\section{The Structure of the Model}

     We imagine well-collimated, relatively narrow, conical, bipolar jets 
that emerged from the supernova core during the explosion event.
The apex of the cones is at the supernova center, but we picture
the jet cones as truncated at the inner and outer ends.
The inner and outer ends we assume are sections of spherical shells
concentric about the center of the ejecta.
This assumption allows us to easily relate jet parameters
and jet mass and kinetic energy as we do in this section and
gives a simple mental picture for our formulation.
In reflection phase of the jets (see \S~3.3), we will, in fact, make
the contradictory assumption that the inner ends are hemispheres with round sides 
toward the observer.
This may be physically plausible and it permits an analytical treatment.
The contradictory assumptions about the inner ends is not a concern.
The structure of the real bipolar jets may be complex and we
make assumptions that allow us to derive simple analytical formulae
that we hope capture the main effect of such jets.

     The conical shape of the jets
seems physically reasonable for material ejected from the
supernova in an axisymmetric fashion as soon as the coasting
phase (homologous phase:  see below) has set in.
We assume the jets are symmetric about the supernova center.
The bulk supernova is assumed to be spherically symmetric.
The bulk supernova and the jets constitute an axisymmetric system
and thus the polarization position angle will be a constant.
Given the nature of electron scattering (see \S~3.3), this position angle
will be perpendicular to the jet axis projected on the sky.
(Position angle shifts of $90^{\circ}$ are possible in general for
axisymmetric systems, but will not occur for the bipolar jet model.
See the discussion of the jet polarization in \S~3.3.) 

     The model as just sketched is not, of course, strictly applicable
to jets in general.
Still one can apply the model to jets in general with the understanding that
the parameter values must be treated as characteristic values, not
well-defined values.
Thus, the model can be applied to observations even if one is
not assuming any precise structure for the jets which may for
instance be more like clumps or filaments.
Simple variations on the model can also developed for analyzing actual
observations.

     We consider only the homologous expansion epoch of the ejecta 
and assume that the jets are participating in the homologous expansion.
In homologous expansion, the radial position $r_{i}$ of any
mass element $i$ is given by
\begin{equation}
r_{i}=v_{i}t=\beta_{i}ct \,\, ,
\end{equation}
where $v_{i}$ is the element's constant velocity, $\beta_{i}$ is
this velocity in units of $c$, and $t$ is the time since explosion.
(At explosion the supernova is usually effectively a point compared 
the size of the ejecta at observable epochs, and so initial radii
are neglected.)
Thus in the homologous expansion
epoch, velocity is a good comoving-frame coordinate
for the ejecta.
In this paper we will usually write velocity in units of $c$:  i.e.,
we use $\beta$'s.

    The inner and outer radial velocities of the jets at their inner and
outer ends are $\beta_{\rm in}$ and 
$\beta_{\rm in}+\Delta\beta_{\rm rad}$, respectively:
$\Delta\beta_{\rm rad}$ is the radial velocity length of a jet.
The velocities  $\beta_{\rm in}$ and $\Delta\beta_{\rm rad}$
are time-independent for the homologous expansion epoch, of course.
The velocity of the supernova photosphere is $\beta_{\rm ph}$:
the photosphere is entirely in the bulk supernova.
The photospheric velocity is time-dependent since the photosphere
recedes into the ejecta as the density and opacity decline with
expansion.
We assume that $\beta_{\rm in}>>\beta_{\rm ph}$:  i.e., the jets
are well detached from the bulk supernova.
This will allow us to approximate the photosphere as a point source.

     The angle of the jet axis from the line-of-sight axis in
the direction of the observer is $\theta$.
We will often represent this angle by its cosine $\mu=\cos\theta$.
We restrict $\theta$ to the range $[0,90^{\circ}]$,
and thus $\mu$ to the range $[0,1]$.
The near jet to the observer is at $\mu$ and the far jet at $-\mu$.

    The half-opening angle of a jet is $\Delta\theta$.
The fractional solid angle $\Omega_{\rm fr}$
(i.e., solid angle divided by $4\pi$) at the supernova center
subtending a jet is given by
\begin{equation}
\Omega_{\rm fr}(\Delta\theta)={1\over2}[1-\cos(\Delta\theta)] 
             \approx {1\over4}\Delta\theta^{2}
             \left(1-{1\over12}\Delta\theta^{2}\right)
             \approx {1\over4}\Delta\theta^{2} \,\, ,
\end{equation}
where the second expression is the 5th order good
approximation to the first and the third is the 3rd-order good
approximation to the first expression.
The 3rd order good expression (used with
$\Delta\theta$ in radians) is accurate to better than 
$\sim 2.3\,$\% for $\Delta\theta\leq 30^{\circ}$
and
$\sim 5.3\,$\% for $\Delta\theta\leq 45^{\circ}$.

    If we assume that the member of the BBK sample of 6 
SNe~Ib that did not conform to the mean photospheric velocity
curve was non-conforming because the line of sight from the observer to 
the supernova center passed through a jet, the estimate of the 
probability of the line of sight passing through a jet is $1/6$. 
(We have no great confidence in this estimate because of the
smallness of the sample and the fact that the bipolar jet model
may not be right.)
If we assume that all bipolar jet supernovae have the same $\Delta\theta$, 
then the estimated probability implies 
that $2\Omega_{\rm fr}(\Delta\theta)\approx 1/6$ or 
$\Delta\theta$ of order $35^{\circ}$.
(The $2$ factor is because our model has two jets.)
We will not, however, take $35^{\circ}$ as
a fiducial half-opening angle $\Delta\theta$:  it is not
at all a definitive result.
The assumptions we will make about the scattering from the
jet in \S~3 are, in fact, better the smaller $\Delta\theta$.

     We will assume we can
treat the jets as geometrically narrow since this allows the
picture of jet scattering we rely on in \S~3.4.
For this assumption to be really true
\begin{equation}
\Delta\theta\beta_{\rm in}<<\Delta\beta_{\rm rad}
\qquad\hbox{or}\qquad\Delta\theta<< b \,\, , 
\end{equation}
where $\Delta\theta$ is in radians, of course, and
\begin{equation}
 b\equiv {\Delta\beta_{\rm rad}\over\beta_{\rm in}} \,\, .
\end{equation}
The parameters $b$ and $\beta_{\rm in}$ are chosen as two of the independent
basic parameters of the model.
The parameter $\Delta\beta_{\rm rad}$ is a dependent parameter with 
this choice.
We do study the behavior of the relative flux and polarization
formulae (see \S~3) for $\Delta\theta$ values 
that are fairly large and cases with $\Delta\theta>b$
in \S~4 with the
understanding that our assumed pictures of scattering
in the jet (see \S~3) are becoming poorer and
must become invalid at some point 
as $\Delta\theta$ grows large or $b$ small.

     We will neglect occultation effects in our model.
Occultation (partial or total) 
will occur, of course, for $\theta$ sufficiently small:
i.e., the near jet will occult the supernova photosphere and
perhaps the far jet;  the supernova photosphere will occult the far jet.
Given our assumption that the jets are well detached 
(i.e., $\beta_{\rm in}>>\beta_{\rm ph}$),
occultation should occur only for relatively small $\theta$:  in the
limit of $\beta_{\rm ph}\to 0$, occultation would only occur
for $\theta\leq\Delta\theta$.
Our expression for polarization will in fact go to zero for
$\theta$ going to zero which is also what it should do if
occultation were considered:  the supernova and jets are
axisymmetric about the line of sight when $\theta=0$, 
and so cannot yield a net polarization whether or not
occultation is considered.
There may well be distinct effects on the line spectra
caused by jet occultation.
But those effects and the occultation situation
in general are beyond the scope of this paper 
and we leave them for future consideration {\it sine~die}.

     We assume the jets have a spatially constant electron density, density
and ionization state.
We also assume the ionization state is constant in time which
means their electron density falls as $t^{-3}$ due to
homologous expansion.
The jet electron density is
\begin{equation}
n_{e}={\rho\over m_{\rm a}\mu_{e}}
     ={M\over  m_{\rm a}\mu_{e} \Omega_{\rm fr}(\Delta\theta)
      (4\pi/3)\left[(b+1)^{3}-1\right]
        \beta_{\rm in}^{3}c^{3}t^{3} }
      \,\, ,
\end{equation}
where $\rho$ is mass density, $m_{\rm a}=1.66053886\times10^{-24}\,$g 
is the atomic mass unit,
$M$ is the total mass of a jet, and
$\mu_{e}$ is the mean atomic mass per electron which is
determined from 
\begin{equation}
{1\over \mu_{e}}=\sum_{i}{X_{i}Z_{i}\over A_{i}} 
\end{equation}
(e.g., \citealt{clayton1983}, p.~84): for ion $i$,
$X_{i}$ is mass fraction, $Z_{i}$ ionization stage,
and $A_{i}$ atomic mass.
Assuming constant ionization state in time implies 
that $\mu_{e}$ is constant in time too. 

    The radial optical depth of a jet is
\begin{eqnarray}
\tau_{\rm rad}&=&n_{e}\sigma_{e}\Delta\beta_{\rm rad}ct  \\
&=&{3\over 4\pi}{\sigma_{e}\over m_{\rm a}}
  {M\over  \mu_{e} \Omega_{\rm fr}(\Delta\theta)}
{b\over\left[(b+1)^{3}-1\right]}
{1\over\beta_{\rm in}^{2}c^{2}t^{2}}
        \nonumber \\ 
&\approx& 2.309\times      
    \left({M\over 0.1\,M_{\odot}}\right)
    \left({4\over \mu_{e}}\right)
    \left({20^{\circ}\over\Delta\theta}\right)^{2}
\left[{(2^{3}-1)b\over (b+1)^{3}-1}\right] \nonumber \\
&&\qquad\qquad
    \left({0.06\over\beta_{\rm in}}\right)^{2}
    \left({20\,\hbox{days}\over t}\right)^{2}
     \,\, , \nonumber 
\end{eqnarray}
where $\sigma_{e}=0.665245873\times10^{-24}\,{\rm cm^{2}}$ 
is the Thomson cross section for electron scattering and
the final expression is in terms of fiducial values with 
$\Omega_{\rm fr}(\Delta)$ approximated by
$\Delta\theta^{2}/4$ (which is accurate to better than
$\sim 2.3\,$\% for $\Delta\theta\leq 30^{\circ}$
and $\sim 5.3\,$\% for $\Delta\theta\leq 45^{\circ}$ as mentioned above).
We emphasize that $\tau_{\rm rad}$ decreases monotonically with time
and becomes constant (in fact 0) only when time $t$ goes to infinity. 

   At present there is insufficient guidance for the fiducial values.
The total mass ejected by typical SNe~Ib is perhaps only $5M_{\odot}$ or less
(e.g., \citealt{hachisu1991}), and one might guess
the bulk spherical symmetry cannot be maintained unless a
jet mass was of order tenths of a solar mass or less:  hence
we choose $0.1\,M_{\odot}$ as the fiducial jet mass.
The ejecta is expected to be helium enriched with intermediate
mass elements, and perhaps some hydrogen:  we have chosen the
fiducial $\mu_{e}$ to be that of singly ionized helium with the
helium atomic mass set to 4 exactly:  thus the fiducial $\mu_{e}=4$.
The fiducial $\Delta\theta=20^{\circ}\approx 1/3\,$radians and $b=1$ are 
just a round numbers consistent with 
$\Delta\theta<<b$ (with $\Delta\theta$ in radians of course)
(i.e., consistent with geometrically narrow jets)
that turn out to yield fairly high polarization (see \S~4.2).
The fiducial time $t=20\,$days was chosen since that is the order of
the rise time to $V$~maximum for low-envelope-mass core-collapse 
supernovae (see \S~1.2).
The fiducial $\beta_{\rm in}=0.06$ 
(i.e., inner end velocity $\sim 18,000\,{\rm km/s}$)
is about twice the
photospheric velocity at $V$ maximum and about three times
the photospheric velocity at 30 days after $V$ maximum
according to the mean photospheric velocity curve of the BBK subsample.
The choice of $0.06$ as the fiducial $\beta_{\rm in}$
is consistent for $V$ maximum epoch (marginally at least) and for later
times with
our assumption that $\beta_{\rm in}>>\beta_{\rm ph}$ which allows
us to approximate the photosphere as a point source.

      We note that with our fiducial values, the jets will have
considerable kinetic energy.
For one jet, a characteristic kinetic energy is
\begin{equation}
{\rm KE}_{\rm char}\approx{1\over2}Mv_{\rm char}^{2}\approx 
      0.724\times\left({M\over 0.1M_{\odot}}\right)
      \left[{\beta_{\rm in}(1+b/2)\over 0.09}\right]^{2}  \, {\rm foe}
      \,\, ,
\end{equation}
where $v_{\rm char}=\beta_{\rm in}(1+b/2)c$ is 
a characteristic velocity, 
foe stands for $10^{51}\,$ergs (the standard unit of
supernova explosion kinetic energies), and we have used the fiducial
values for $M$, $\beta_{\rm in}$, and $b$ (see eq.~[7] above).
Since the whole kinetic energy of an ordinary supernova
is of order $10^{51}\,$ergs 
(e.g., \citet{woosley1994}, p.~90), our fiducial values
and $v_{\rm char}$ value require
that the jets make large energy demands on the explosion mechanism.
We are not concerned about this here.
First, our fiducial values and $v_{\rm char}$ value were 
chosen without much guidance:
perhaps less demanding values will turn out to be better
in actual polarization spectrum analyses.
Second, the actual energy in neutrinos available in 
core-collapse explosions is of order $10^{53}\,$ergs
(e.g., \citet{woosley1994}, p.~90).
It is thought that some of this available energy is transformed
into the kinetic energy of the ejecta and is the main energy
of the explosion. 
But how this is done exactly 
and how much energy is transformed are still very uncertain
(e.g., \citealt{janka2004}).
Thus, we are plausibly free to invoke jet energies of order foes
for our hypothetical jets.
Hydrodynamic jet calculations also somewhat freely invoke 
jet energies 
(e.g, \citealt{hoeflich2001, macfadyen2001, zhang2003}).

     For our choice of fiducial values, we obtain 
jet radial optical depth $2.309$ which is sufficient 
to produce fairly large polarization:  see \S~4.2.
However, as equation~(7) shows the optical depth decreases
as $1/t^{2}$.
Thus for increasing time with the other parameters fixed the
optical depth falls and eventually polarization must decrease
because few photons are being scattered and polarized.
For example by day~100 (after the explosion) 
the radial optical is $0.09$ from equation~(7).
Such a low optical depth would usually give very low polarization
compared to earlier times.
But significant polarization for days 100 to 200
can be obtained from a different choice of parameters for the jets.
The plausibility of these choices needs to be investigated.

    In an analysis where one does not want to specify
the jet mass or mean atomic mass per electron $\mu_{e}$,
one can just specify
\begin{equation}
\tau_{\rm rad}=\tau_{\rm rad,0}\left({t_{0}\over t}\right)^{2} \,\, ,
\end{equation}
where $\tau_{\rm rad,0}$ is a fiducial radial optical depth 
at a fiducial time $t_{0}$.
Actually there is only one parameter $\tau_{\rm rad,0}t_{0}^{2}$,
but splitting $\tau_{\rm rad,0}t_{0}^{2}$ 
into two factors is an intelligible reparameterization:
one might chose $\tau_{\rm rad,0}=1$ in all cases and then
$t_{0}$ would become the single parameter.
In this paper, we will often just use $\tau_{\rm rad}$ itself
as a parameter and for time just a reduced time defined by
\begin{equation}
t_{\rm red}={ 1\over \sqrt{\tau_{\rm rad}} }
\qquad\hbox{with the inverse relation being}\qquad
\tau_{\rm rad}={1\over t_{\rm red}^{2}} \,\, .
\end{equation}
Note that when $\tau_{\rm rad}=1$, $t_{\rm red}=1$.
The $t_{\rm red}=1$ epoch is important for jet polarization.
Once the optical depth through a jet falls below $\sim 1$,
the fraction of direct flux being scattered by the jet 
must start
declining, and thus so must the jet polarization (see \S~4.2).

    Using equations~(7) and~(10) we find the 
relation of time to reduced time to be 
\begin{eqnarray}
t&=&t_{\rm red}\sqrt{\tau_{\rm rad,0}t_{0}^{2}}
  =t_{\rm red}
\sqrt{ {3\over 4\pi}{\sigma\over m_{\rm a}}
      {M\over  \mu_{e} \Omega_{\rm fr}(\Delta\theta)}
  {b\over\left[(b+1)^{3}-1\right]}      }
  {1\over\beta_{\rm in}c}  \\
&\approx& 30.39\,\hbox{days}\quad\times\quad t_{\rm red}
  \sqrt{ \left({M\over 0.1\,M_{\odot}}\right)
         \left({4\over \mu_{e}}\right)
         \left[{(2^{3}-1)b\over (b+1)^{3}-1}\right] }
\left({20^{\circ}\over\Delta\theta}\right)
       \left({0.06\over\beta_{\rm in}}\right)
			           \,\, ,  
\end{eqnarray}
where the final expression is in terms of our fiducial values for
the parameters
and we have approximated $\Omega_{\rm fr}(\Delta)$ by $\Delta\theta^{2}/4$
again.
If the fiducial values are used, we see that $t_{\rm red}=1$ and 
$\tau_{\rm rad}=1$ occur at day 30.39 after the explosion.

    Another quantity we will need is the radial velocity,
which we will call $\beta_{\rm scat}$, which locates
the place in the jet that is the mean optical depth into the jet that
photons radially incident on the jet reach for their first scattering
under the assumption that they do scatter in the jet.
This mean optical depth $\tau_{\rm mean}$ is given by
\begin{equation}
\tau_{\rm mean}=\cases{\displaystyle 
  1-{\tau_{\rm rad}\over
     \exp(\tau_{\rm rad})-1}\,\, & in general;  \cr
     1\,\,                       & for $\tau_{\rm rad}>>1$;  \cr
\noalign{\medskip}
\displaystyle {1\over2}\tau_{\rm rad}
 \left(1-{\tau_{\rm rad} \over 6}
         +{\tau_{\rm rad}^{3} \over 360} \right) \,\,  
                    & for $\tau_{\rm rad}\lesssim 1$.  \cr}
\end{equation}
At $\tau_{\rm rad}=0$, $\tau_{\rm mean}=0$;  
$\tau_{\rm mean}$ then grows
monotonically with $\tau_{\rm rad}$ initially with slope $1/2$;
but the slope decreases monotonically with $\tau_{\rm rad}$ 
and $\tau_{\rm mean}$ goes asymptotically to $1$ as
$\tau_{\rm rad}\to\infty$:  
$\tau_{\rm mean}(\tau_{\rm rad}=1)\approx0.418$,
$\tau_{\rm mean}(\tau_{\rm rad}=3)\approx0.843$,
and
$\tau_{\rm mean}(\tau_{\rm rad}=5)\approx0.966$.
The only stationary point of $\tau_{\rm mean}$ is a maximum at 
$\tau_{\rm rad}=\infty$.
In supernova time evolution, 
$\tau_{\rm mean}$ is nearly a constant $1$ at early times
when $\tau_{\rm rad}$ is large, and then
decreases as $\tau_{\rm rad}$ decreases;
$\tau_{\rm mean}$ is decreasing roughly linearly with $\tau_{\rm rad}$ when 
$\tau_{\rm rad}$ falls to of order 1;  
then $\tau_{\rm mean}$ vanishes linearly with 
$\tau_{\rm rad}$ as time goes to infinity.

     Since the electron density is assumed uniform,
$\beta_{\rm scat}$ is given by 
\begin{equation}
\beta_{\rm scat}
=\beta_{\rm in}+\Delta\beta_{\rm rad}
		 {\tau_{\rm mean}\over\tau_{\rm rad}}
=\Delta\beta_{\rm rad}\left(b^{-1}
     + {\tau_{\rm mean}\over\tau_{\rm rad}}\right) 
=b\beta_{\rm in}\left(b^{-1}
     + {\tau_{\rm mean}\over\tau_{\rm rad}}\right) \,\, .
\end{equation}
We take $\beta_{\rm scat}$ to be the characteristic
comoving-frame radius for scattering in the jet.
At late times when $\tau_{\rm rad}\to 0$,
$\beta_{\rm scat}\to \beta_{\rm in}+\Delta\beta_{\rm rad}/2$.

    We will also need 
the optical depth perpendicular to the jet axis from the
jet axis to the jet surface at $\beta_{\rm scat}$.
We call this the off-radial optical depth $\tau_{\rm off}$.
The approximate expression for $\tau_{\rm off}$ (with $\Delta\theta$ in
radians) is
\begin{equation}
\tau_{\rm off}=\Delta\theta \beta_{\rm scat}
               {\tau_{\rm rad}\over\Delta\beta_{\rm rad}}
	      =\Delta\theta\left({\tau_{\rm rad}\over b}
	       +\tau_{\rm mean}\right) \,\, .
\end{equation}
For high polarization, it is good to have $\tau_{\rm rad}$ large for
lots of polarizing scattering and $\tau_{\rm off}$ small for
low depolarizing multiple scattering (see \S~3.4).
This means $\Delta\theta/b<<1$ is a desirable condition for
high polarization:
we have, in fact, already indicated our need for this condition (which
implies a geometrically narrow jet) above (see eq.~[3]).
In another perspective, $\tau_{\rm rad}$ large and  $\tau_{\rm off}$ 
small together
imply a geometrically narrow jet since we are assuming constant
electron density in the jet.

    Mathematically, $\tau_{\rm off}$ always decreases with decreasing
$\tau_{\rm rad}$ or increasing time.
Its behavior is summarized in the following expressions:
\begin{equation}
\tau_{\rm off}\approx\Delta\theta\cases{
\displaystyle  \left({\tau_{\rm rad}\over b}
                  +\tau_{\rm mean}\right)  & exactly; \cr 
\noalign{\smallskip}
\displaystyle {\tau_{\rm rad}\over b} \,\,  & for early times when
				         $\tau_{\rm rad}>>1$ and
					 $\tau_{\rm rad}>>b$; \cr
\noalign{\smallskip}
\displaystyle {\tau_{\rm rad}\over b} \,\,  & for all times if $b<<2$; \cr
\noalign{\smallskip}
\displaystyle \tau_{\rm rad}\left({1\over b}+{1\over2}\right) \,\,  
                          & for late
                          times when
      	                  $\tau_{\rm rad}\lesssim 1$; \cr
\noalign{\smallskip}
\displaystyle 1 \,\,   & for intermediate times if $b>>2$;  \cr
                       & $\tau_{\rm rad}/b<<1$, and $\tau_{\rm rad}>>1$. \cr}
\end{equation}
Note the following also.
If $b<2$, then ${\tau_{\rm rad}/b}>\tau_{\rm mean}$ always, except at
time going to infinity where both $\tau_{\rm rad}$ and $\tau_{\rm mean}$
are zero.
If $b=2$, then ${\tau_{\rm rad}/b}$ and $\tau_{\rm mean}$ asymptotically
approach each other as time goes to infinity. 
If $b>2$, then there is a phase where ${\tau_{\rm rad}/b}<\tau_{\rm mean}$,
that ends only when time goes to infinity.

      The time evolution of $\tau_{\rm rad}$ and $\tau_{\rm off}$
control the time evolution of bipolar jet model polarization as we
will see from the analytic polarization formula equation~(48) 
and equation~(42) in \S~3.5. 
An interesting prediction of the bipolar jet model emerges 
from this control is
there are two kinds of phases in which the continuum polarization
can plateau for some time interval if we assume that no
significant time dependence affects the continuum polarization
through the redshift effect (see \S~3.3). 
The first kind occurs at early times going back to time zero when
$\tau_{\rm rad}$ and $\tau_{\rm off}$ are both effectively infinite
for a finite time:  effectively infinite since they enter the
function controlling polarization evolution through the
functions $1-\exp(-\tau_{\rm rad})$ and $\exp(-\tau_{\rm off})$
(see \S~3.5, eq.~[42] and~[48]) and make those functions nearly
exactly 1 and 0, respectively for a finite time.
This kind of polarization plateau
phase is demonstrated in the calculations in \S~4.2:
see the early-time, non-infinite $b$ polarization evolution curves
in Figures~1, 2, and~3 in \S~4.2.

    The second kind of polarization plateau phase occurs 
at intermediate times for $b>>2$ or back to time zero for $b=\infty$.
In this case, $\tau_{\rm rad}$ is effectively infinite in
the functions $1-\exp(-\tau_{\rm rad})$ and
$\tau_{\rm mean}$ making both these functions nearly 1
for a finite time, but, on the other hand, 
${\tau_{\rm rad}/b}<<\tau_{\rm mean}\approx 1$ for a finite
time (which situation requires $b>>2$) and this makes
$\tau_{\rm off}\approx\Delta\theta$ (with $\Delta\theta$
in radians, of course) for a finite time. 
In other words, one requires ${\tau_{\rm rad}/b}<<1$ while
$\tau_{\rm rad}>>1$ for the second kind of polarization plateau. 
The second of kind of polarization plateau phase is also
demonstrated in \S~4.2:
see the polarization evolution curves with $b\gtrsim 100$ in 
Figures~1 and~3.

    That the polarization plateau 
phases can exist is a mathematical prediction of the bipolar jet model.
The first kind of plateau phase is unavoidable if our assumptions
are adequate and should be observable
if a supernova with jets is observed early enough.
The second kind can only exist if $b>>2$:  we show calculations
without the second kind of polarization plateau
in in Figures~2 and~3 in \S~4.2.
The second kind of polarization plateau is probably physically
unrealizable for cases of significant polarization as we
discuss in \S~4.2.

\section{The Radiative Transfer of the Model}

     In the derivation of the bipolar jet model formulae
for scattered flux, relative scattered flux,
and polarization, we make many simplifying assumptions.
These assumptions are needed to capture the main effect of the 
model on flux and polarization analytically and tractably.
One main simplifying assumption is that
the only significant opacity in a jet
is electron scattering (or Thomson) opacity.
Thus we neglect the scattering, absorbing, and
depolarizing effects of other opacities of which
line opacity is probably the most important.
Another consequence of assuming only electron
scattering opacity is that no photons are destroyed in the jet.
We also assume none are created.

      In \S~3.1, we will make the point source assumption for
the spherically symmetric bulk supernova.
Subsection~3.2, gives a simplified treatment of the time-delay
effect which could be incorporated into the bipolar jet model.
In the following three subsections, we derive analytic formulae 
for scattered flux, scattered relative flux, 
and polarization spectra that account
for the three phases of jet polarization.
The first phase, which we call the reflection phase,
is when the jet is optically thick in all directions (see \S~3.3).
The second phase, which we call the radial phase,
is when jet is optically thick in the radial direction, but
optically thin in the direction perpendicular to the jet
axis (see~3.4).
The third phase, which we call the optically-thin phase,
is when the jet is optically thin in all directions (see \S~3.4).
In \S~3.5, we give general expressions that are for all phases.

\subsection{The Point Source Approximation for the Supernova}

     Given that the direct luminosity from the supernova
(i.e., the luminosity without the jet contribution)
at rest-frame wavelength $\lambda$ is 
$L_{\rm dir}(\lambda)$,
the direct flux at Earth from the supernova (i.e., the flux
without the jet contribution) is
\begin{equation}
f_{\rm dir}(\lambda)={L_{\rm dir}(\lambda)\over 4\pi D^{2}} \,\, ,
\end{equation}
where $D$ is the Earth-supernova-center distance.

     The fraction of $L_{\rm dir}(\lambda)$ that enters the
inner end of a jet is just $\Omega_{\rm fr}(\Delta\theta)$ (see eq.~[2]
in \S~2).
This is true regardless of the size of the supernova photosphere
(provided $\beta_{\rm ph}\leq\beta_{\rm in}$):
it is a consequence of the assumed spherical symmetry of the 
spherical bulk of the supernova.
We will assume that $L_{\rm dir}(\lambda)\Omega_{\rm fr}(\Delta\theta)$ is all
the luminosity that passes through a jet and that all the beams
from the bulk supernova
are radial:  i.e., all the beams from the bulk supernova
that enter a jet are radial and they
enter at the inner end and they leave at the outer end.
Thus we are making the point-source approximation for the
photosphere.
This approximation should be reasonably good as long as
$\beta_{\rm in}>>\beta_{\rm ph}$ as we have already assumed
(see \S~2).
There are, of course, two errors associated with the approximation caused
by non-radial beams from a finite photosphere.
First, there will be non-radial beams  
entering the inner end of a jet (these are counted in 
$L_{\rm dir}(\lambda)\Omega_{\rm fr}(\Delta\theta)$, of course) 
and these beams will not always traverse the whole radial extend of the jet.
Second, there will be non-radial beams that enter the sides
of a jet (these are not counted in    
$L_{\rm dir}(\lambda)\Omega_{\rm fr}(\Delta\theta)$, of course) and these beams
will only traverse some part of the radial extent of the jet.
The first error leads to an overestimate of 
and the second error, to an underestimate of the direct-flux-jet interaction
when we make the point-source approximation.
The two errors are somewhat compensating:  this somewhat
strengthens point-source approximation.
The point-source approximation is also somewhat strengthened
by the fact that the photospheric emission will be somewhat
limb-darkened:  thus the photosphere will be more point-like
than assuming that angle-independent specific intensity
beams emerge from it.
Thus, our photosphere is more point-like than a non-zero 
$\beta_{\rm ph}$ suggests at first.

\subsection{The Time-Delay Effect}

    The time-delay effect is caused by the supernova evolving during the
time that flux travels from the photosphere to the jets and during
the radiative transfer through the jets.
In this section we will develop some formalism to treat the time-delay
effect in the bipolar jet model, but we will not actually incorporate this
formalism into our formulae in \S\S~3.3, 3.4, and 3.5 and
our calculations in \S~4.
The variations on those formulae that allow for the time-delay
effect can easily be derived if needed, but are a bit cumbersome
in appearance.

    In general, time-delay effect will be complex.
So we will make some simplifying assumptions.
We will assume again that the bulk supernova can be treated
as a point at the center of the system and we will assume that
the jet can be treated as a point at 
velocity $\beta_{\rm scat}$ (see \S~2, eq.~[14]) 
for time-delay effect purposes:  this eliminates
the need to consider the time delays while flux transfers through
the jets.
Now $\beta_{\rm scat}$ itself is time dependent.
As a simplification we assume it can be evaluated at 
a fiducial time $t$:  
$t$ and all other times below are counted from 
the time of the supernova explosion:  i.e., since
time zero.
This last assumption is actually rather poor since
the time-delay effect will cause us to evaluate 
jet optical depth parameters at times different than $t$
(as we will see below) and those optical depth parameters
determine $\beta_{\rm scat}$.
But to get a sense of the time-delay effect, it is adequate.

    For each time of observation of flux and polarization spectra, 
we will in general have to consider 5 supernova evolution times.
But we will compact the expressions by using an upper case label for
the near jet and a lower case label for the far jet.
Let $t_{\rm jet\pm}$ be the time that flux arrives at a jet 
from the bulk supernova.
Let $t_{\rm jet\pm}'$ be the time that same flux started out from the 
bulk supernova.
The direct flux from the bulk supernova that starts at fiducial 
time $t$ arrives at the Earth at the same time as the flux
from the jet that starts at time $t_{\rm jet\pm}$:  as
mentioned above, we assume no time delay for the flux inside the jet. 
From geometry and homologous expansion, it follows that
\begin{equation}
 t_{\rm jet\pm}=t_{\rm jet\pm}'+{\beta_{\rm scat}ct_{\rm jet\pm} \over c}
 \qquad\hbox{and}\qquad
 t_{\rm jet\pm}=t\pm{\mu\beta_{\rm scat}ct_{\rm jet\pm} \over c} \,\, ,
\end{equation}
where $\mu$ (as specified in \S~2) is the cosine of the angle to the near 
jet:  we confine $\mu$ to the range $[0,1]$.
A little algebra leads to
\begin{equation}
t_{\rm jet\pm}'=t{1-\beta_{\rm scat}\over 1 \mp \mu\beta_{\rm scat} }
\qquad\hbox{and}\qquad
t_{\rm jet\pm}=t{1\over 1 \mp \mu\beta_{\rm scat} }  \,\, .
\end{equation}

    From the expressions of equations~(18) and~(19), 
we have at once the inequalities
\begin{equation}
t_{\rm jet+}'\leq t\leq t_{\rm jet+} 
\qquad\hbox{and}\qquad
t_{\rm jet-}'\leq t_{\rm jet-}\leq t  \,\,  .
\end{equation}
The inequalities show the time ordering of the events.
Note that $t_{\rm jet\pm}=t$ only when $\mu\beta_{\rm scat}=0$
and $t_{\rm jet\pm}'=t_{\rm jet\pm}$ only when $\beta_{\rm scat}=0$.
Note also that the ratios of the time parameters are not constant since
$\beta_{\rm scat}$ is itself time dependent.
We have recall simplified this time dependence by assuming
that we can evaluate $\beta_{\rm scat}$ at the fiducial time
$t$ itself.

    Equation~(19) is the probably the most useful time-delay effect
result since fiducial time $t$ is the natural time for any calculation:  
it is the time that characterizes the bulk of
the flux that ones sees:  sees $D/c$ later than $t$, of course:
recall $D$ is the Earth-supernova-center distance.

    As we will see in \S\S~3.3, 3.4, and 3.5, polarization is high
for $\mu\approx0$. 
For example, for a high polarizing jet one might have $\mu=0$
and $\beta_{\rm scat}=0.1$:  this implies $t_{\rm jet\pm}'=0.9t$.
Thus, with these parameters the observer detects a mixture
of direct flux and scattered flux where the scattered flux
originated in the bulk supernova
at at time $10\,$\% earlier than the time of origin of the direct flux.
Since considerable evolution in flux level and spectra can occur
in $10\,$\% of the time since explosion, the time-delay effect could
be significant.
Thus, in general it would be useful to have a
complete enough time coverage of observed flux spectra so that one can
use the observed spectra from earlier times to construct synthetic
polarization spectra using the bipolar jet model for later times.

     Equation~(19) can be used with
reduced times (see \S~2) replacing ordinary times:  i.e.,
$t$ becomes $t_{\rm red}$, $t_{\rm jet\pm}$ becomes $t_{\rm red\,\,jet\pm}$,
and $t_{\rm jet\pm}'$ becomes $t_{\rm red\,\,jet\pm}'$.
Using $t_{\rm red}$ as the fiducial time and equation~(19), and
equations~(10), (13), and~(15) of \S~2, one can calculate the 
needed time-delay affected optical depths for $t_{\rm red}$.
Also using $t_{\rm red}$ and equation~(19), one can 
calculate the times at which the flux from the jets started out
from the supernova (i.e., $t_{\rm red\,\,jet\pm}'$), and 
thus when calculating the net flux one
can add the direct and jet fluxes corrected for the time delay
effect. 
In fact, $t_{\rm red}$ (which replaces $t$) can be used
as of the 5 independent basic parameters of the bipolar jet model 
when the time-delay effect is included.
If one neglects the time-delay effect, $\tau_{\rm rad}$ is the same
for both jets in a calculation and $\tau_{\rm rad}$ can be used
instead of  $t_{\rm red}$ as one of the 5 independent basic parameters.

    We will not explore the consequences of the time-delay effect 
in detail in this paper.
But we will briefly discuss them here.
As a heuristic formula (suggested by eq.~[48] in \S~3.5), 
let polarization as a function of $t$ be given for the nonce by 
\begin{equation}
P(t)\sim P_{0}(t)\Pi(t) \,\, ,
\end{equation}
where $P_{0}(t)$ is polarization in the absence of time delay and
\begin{equation}
\Pi(t)={f_{\rm dir}(\bar t_{\rm jet}')\over f_{\rm dir}(t)} \,\, ,
\end{equation}
where $f_{\rm dir}$ is the direct flux from the bulk supernova and
$\bar t_{\rm jet}'$ is some appropriate average of $t_{\rm jet+}'$
and $t_{\rm jet-}'$.
As shown by the demonstration results in \S~4, there will be a 
time of maximum (continuum) polarization.
These demonstration calculations were done neglecting the time-delay effect,
and so their polarization corresponds to what we have labeled for
the nonce $P_{0}(t)$.
Clearly, if $\Pi(t)$ is a rising/falling function at the time of the
$P_{0}(t)$ maximum (which is a stationary point recall), 
then the $P(t)$ maximum will be shifted to
a later/earlier time than that of $P_{0}(t)$ maximum.

    Now it is difficult to predict the temporal behavior of $\Pi(t)$ in
general since it depends on the shape of the supernova light curve
and on the how $\bar t_{\rm jet}'$ varies with time.
But near maximum light, $\Pi(t)$ should be a rising function
since the $f_{\rm dir}(\bar t_{\rm jet}')$ will be rising and
$f_{\rm dir}(t)$ should be nearly stationary.
So if the $P_{0}(t)$ maximum is at about maximum light,
the time-delay effect will tend to give the $P(t)$ maximum at a later time.

   At late times (i.e., probably of order 100 days or more after explosion), 
$\beta_{\rm scat}$ will approach a constant $\Delta\beta_{\rm rad}(b^{-1}+1/2)$
as $\tau_{\rm rad}\to0$ (see eq.~[13] and~[14] in \S~2),
and thus the ratio $\bar t_{\rm jet}'/t$ will approach a constant too
(see eq.~[19]):
a constant that is
less than 1, of course (see eq.~[20]).
At such times the light curve of the supernova is falling exponentially or 
quasi-exponentially (e.g., \citealt{jeffery1999}), and
so $\Pi(t)\approx \exp[(1-\bar t_{\rm jet}'/t)t/t_{e}]$ which is 
growing exponentially:  $t_{e}$ is the assumed $e$-folding time of the
light curve and the ratio $\bar t_{\rm jet}'/t$ is a constant less
than or equal to 1 (see eq.~[19] and [20]).
Thus if the $P_{0}(t)$ maximum occurs in this late phase, the $P(t)$ maximum
will tend to occur at a later time again. 
Actually, because ejecta is tending to be optically thin in the
exponential/quasi-exponential phase, one does not expect 
the $P_{0}(t)$ maximum to occur then. 

    In the analytic formulae we derive in the remainder of this
section and in the calculations reported in \S~4, 
we will, as mentioned above, neglect the time-delay effect.
This means we approximate all the time parameters
by one value, $t$ or $t_{\rm red}$.
When only one time parameter is needed then, as mentioned above, 
$\tau_{\rm rad}$ can be used instead
of $t_{\rm red}$ as one of the 5 independent basic parameters.

\subsection{Reflection Phase}

    In the reflection phase, the jet is optically thick and 
it acts as a diffuse reflector of incident light.
Recall that for simplicity we have assumed that the jets have uniform
electron density.
In fact they are likely to increase in electron density as
one moves inward.
Thus photons striking the optically thick jet that penetrate
somewhat deeply in optical depth will scatter
a few times and then tend to diffuse outward again and be
emitted from the jet.
The behavior of these deep-penetration photons is effectively uninfluenced
by their history prior to entering the jet.
The deep-penetration photons will have an emergent angular
distribution that is rather isotropic over solid angle
$2\pi$ (i.e., the half of the sky not filled by the jet)
and is mostly unpolarized---individual photons will be polarized, but
their net flux mostly not. 
Some anisotropy and net flux polarization will occur due to diffusive
photon scattering in an inhomogeneous medium 
(e.g., \citeauthor{chandrasekhar1960} [1960, p.~248] for plane-parallel
atmospheres;  \citeauthor{cassinelli1971} [1971] for spherically symmetric
atmospheres), but this is likely to be a relatively minor effect 
and too complicated for us to contemplate in the simple picture
we are developing.
There is also a class of photons that scatter more than once,
but not sufficiently to lose all trace of their history
before they entered the jet.
Their angular distribution and polarization state are
still somewhat determined by their initial direction
of incidence on jet:  but this non-random redistribution
and polarization is likely to have a relatively small effect.
We will assume that photons that scatter more than once
in the jet emerge unpolarized and are isotropically
distributed over $2\pi$.

     The photons that scatter once and escape
the jet will have an angular distribution and polarization
determined by the Rayleigh phase matrix
(e.g., see \citeauthor{chandrasekhar1960} 1960, p.~37).
We will estimate the fraction $q$ of photons in this
category in the following way.
The probability that a photon penetrates a jet along some
typical beam path to optical depth $\tau$ without scattering
and then scatters in optical depth $d\tau$ is
$\exp(-\tau)\,d\tau$.
Since we are considering a very optically thick phase,
we can regard the local jet surface as a plane-parallel
atmosphere.
Thus $\sim 1/2$ the scattered photons are aimed at a path
toward the ``plane-parallel'' surface and $\sim 1/2$ are directed
inward and must scatter more than one time before escaping:
we assume these fractions are both exactly $1/2$.
This half-inward-half-outward division among scattered photons
is true even with the anisotropic Rayleigh distribution 
since this distribution is axisymmetric about the incident
direction and forward-back symmetric:  any plane through
a polar plot of the distribution slices it into two regions
with the same integrated emission.
We will assume that the average optical depth distance to 
escape the jet for a photon that has just scattered for the
first time is equal to $\tau$:  note the penetrating and
escaping directions are not the same;  we are just assuming
the optical depth along them is the same in some kind of average sense.
Thus the probability of this photon escaping without scattering
again is estimated to be $\exp(-\tau)$. 
If we multiply the relevant factors together, 
we find the estimate of the probability
of a photon incident on the jet penetrating to optical depth
region $d\tau$ and then escaping the jet is
$(1/2)\exp(-2\tau)\,d\tau$.
The estimated $q$ fraction is the integral of 
$(1/2)\exp(-2\tau)\,d\tau$ over all $\tau$:
\begin{equation}
       q\approx{1\over2}\int_{0}^{\infty} \exp(-2\tau)\,d\tau
       ={1\over4} \,\, , 
\end{equation}
where the integral can be extended to infinite optical depth because
we are assuming the jet is optically thick.
If one had a well specified jet structure, one could calculate
more exactly the average fraction of once-scattered photons.
That calculation is too elaborate for our crude picture of the jet.
In this paper, we adopt $1/4$ as the fixed value for $q$.

     To treat polarization we must label polarization axes
for the scattered light.
We will label these axes
(which are both perpendicular to the scattered light's
propagation direction) $x$ and $y$:  the $x$ axis
is perpendicular to the plane of scattering and the $y$ axis
is aligned with it.
The plane of scattering (or scattering plane) is the plane defined 
by incident and scattered beams.
The Stokes $Q$ and $U$ parameters are needed 
to evaluate the polarization:  see  
see \citeauthor{chandrasekhar1960} (1960, p.~25) for 
a detailed presentation of the Stokes parameters.
The Stokes $Q$ parameter for the scattered flux in our
setup is $y$-flux minus the $x$-flux following conventions
of Chandrasekhar \citep[see][p. 27]{chandrasekhar1960}.

     Because of the finite size of the jet there will be a range of 
scattering planes for light scattered in the observer direction.
In our mental picture, we will approximate this range of planes by the 
plane defined by the jet axis and the scattering direction to the
observer:  we will call this plane the fiducial plane.
Just using the fiducial plane alone would overestimate the polarizing 
effect since the polarization-canceling effect from a range of planes 
is neglected.
We will make a crude correction factor for the Stokes $Q$ formula 
to account for the range of scattering planes.

    Consider a circular arc of material in the jet
with arc center at the center of the supernova
in the plane defined by the jet axis and the line
perpendicular to the line of sight.
The arc is meant to approximate the layer of scattering material
in the jet at $\beta_{\rm scat}$.
The angle at the jet center subtended by the arc is naturally $2\Delta\theta$.
Measure the angle for any point along the arc by $\zeta$:
$\zeta=0$ at the jet axis and is positive counterclockwise.
Suppose again that the supernova photosphere is a point.
Emission from the supernova center is scattered off electrons in
the arc toward the observer.
Assume the incident specific intensity on the arc, the amount of
scattering, and the polarizing effect are constant along the arc.
The local Stokes axes at each point along the arc are labeled
$x'$ (tangent to the arc) and $y'$ (perpendicular to the arc).
The unadorned $x$ and $y$ axis are fiducial axes for the fiducial plane.
The local Stokes parameters are $dQ'$ and $dU'$:  $dQ'<0$ from the 
Rayleigh phase matrix (e.g., see \citeauthor{chandrasekhar1960} 1960, p.~37)
and $dU'=0$ by symmetry:
we are assuming the incident light is unpolarized.
The local position angle of polarization is aligned with the $x'$ axis:
see the Rayleigh phase function (eq.~[28]) and accompanying discussion below.
If we could legitimately just integrate up the local $dQ'$ without
transforming to the fiducial $x$-$y$ axes,
then the net $Q'$ for the arc emission would be proportional to
$2\Delta\theta$.
More correctly we should transform $dQ'$ and $dU'$ to the fiducial 
$x$-$y$ axes and then integrate.
The transformations for a rotation $\zeta$ clockwise are 
\begin{equation}
dQ=dQ'\,\cos(2\zeta)\qquad\hbox{and}\qquad
dU=-dQ'\,\sin(2\zeta)
\end{equation}
(\citeauthor{chandrasekhar1960} 1960, p.~34).

   The integration for net $U$ gives zero because the integration interval
is even:  from a more obvious perspective, 
the axisymmetry of the jet requires the net $U$ for the
fiducial plane Stokes axes to be zero.
Because $U=0$, the position angle of polarization is determined only
by the net $Q$.
Given the nature of electron scattering (which we discuss below), the
polarization is aligned perpendicular to the fiducial plane.

   The net $Q$ corrected for the variation of the polarization axes along
the arc in terms of the net $Q'$ neglecting this variation is given by
\begin{equation}
Q=Q'{1\over2\Delta\theta}\int_{-\Delta\theta}^{\Delta\theta} 
    \cos(2\zeta)\,d\zeta
=Q'{1\over\Delta\theta}\int_{0}^{\Delta\theta} 
=Q'{\sin(2\Delta\theta)\over 2\Delta\theta}  \,\, .
\end{equation}
Thus to crudely correct the Stokes $Q$ parameter in our formulae
developed assuming only fiducial plane scattering, we will multiply it by 
the correction factor
\begin{equation}
\xi(\Delta\theta)={\sin(2\Delta\theta)\over 2\Delta\theta}
\approx 1-{1\over6}(2\Delta\theta)^{2}
+{1\over120}(2\Delta\theta)^{4} \,\, ,
\end{equation}
where the second expression is the 5th order good approximation in
small $2\Delta\theta$ (in radians) to the first.
Since the 1st order term in the correction factor vanishes, we would have
1st order good expression for $Q$ without using the correction factor. 
The correction factor is, of course, only a crude correction since 
the jet is not a single arc at a single inclination to the line of sight.
But the $\xi(\Delta\theta)$ factor does crudely have the right effect and
the exactly right limiting behavior. 
In the vanishingly-narrow-jet limit $\Delta\theta\to0$
and $\xi(\Delta\theta)\to 1$ as it should since there is no
range of scattering planes in this case.
In the limit where the jets cover the sky spherically symmetrically
$\Delta\theta\to\pi/2$ and $\xi(\Delta\theta)\to 0$:  this is also
correct since the symmetry would cancel all polarization. 
Between the limits, $\xi(\Delta\theta)$ decreases monotonically with
$\Delta\theta$ with the only stationary point being a maximum at
$\Delta\theta=0$.
At worst, $\xi(\Delta\theta)$ is an adequate interpolation formula
for a correction factor;  it should be somewhat better than that though.

     A similar Stokes $Q$ parameter correction factor
could perhaps be devised to account for the finite size of the supernova photosphere.
But it would complicate our mental picture and, given the crudity
of the model, seems of little value.

     The angular redistribution function for electron
scattering of unpolarized light is the normalized Rayleigh phase function:
\begin{equation}
   g_{\rm norm}(\mu_{\rm gen})={1\over4\pi}g(\mu_{\rm gen}) \,\, \hbox{:}
\end{equation}
$g(\mu_{\rm gen})$ is the (conventional) Rayleigh phase function given by
\begin{equation}
  g(\mu_{\rm gen})={3\over4}\left(1+\mu_{\rm gen}^{2}\right) \,\, ,
\end{equation}
where $\mu_{\rm gen}$ is the cosine of a general
scattering angle (i.e., the angle
between incident and scattered beam directions)
(e.g., \citeauthor{chandrasekhar1960} 1960, p.~35).
The net direct supernova light incident on the jets is unpolarized.
Individual beams are polarized since the supernova atmosphere is 
electron-scattering dominated, but spherical symmetry of the bulk supernova
makes the net polarization zero.
In our notation the $1$ in the $1+\mu_{\rm gen}^{2}$ factor
determines the intensity perpendicular to the scattering plane
(i.e., the $x$-flux)
and the $\mu_{\rm gen}^{2}$, the intensity parallel to the scattering plane
(i.e., the $y$-flux):
this follows from the Rayleigh phase matrix (e.g., 
\citeauthor{chandrasekhar1960} 1960, p.~37).
Note that forward and backward scattering are unpolarized since
$\mu_{\rm gen}^{2}=1$ in those cases and the two scattered flux components
are equal.
Scattering through $90^{\circ}$ is gives the maximum polarization of 
$100\,$\%.

     The Rayleigh phase function is adequate to describe photon
redistribution for us for two reasons.
First, we can assume electron scattering is coherent 
(i.e., does not change wavelength) in the jet frame.
The coherent scattering approximation is adequate since
the wavelength part of the photon redistribution (which
affects the angular redistribution in general) on electron
scattering for electrons obeying a Maxwellian distribution
(\citeauthor{mihalas1978} 1978, p.~420) averages away 
for continuum radiative transfer which is all we are considering in the jet.
Second, the flux in the jet can be approximated as unpolarized in
an average sense before a last scattering which may or may not be
polarizing.
For the reflection phase considered in this section the last scattering
in the jet is only polarizing on average if it is also the first
scattering as we argued above.
For the radial and optically-thin phases of the jet (see \S~3.4),
the photons that are only scattering radially inward and outward
(i.e., radially directed photons)
are approximately unpolarized since there scattering is approximately
backward and forward scattering. 
The last scattering is polarizing if it is of radially-directed, 
and so unpolarized, photon.
If the last scattering of non-radially-directed photon is not polarizing
on average as we will argue in \S~3.4.

    We assume here and in all our developments that all scattering toward 
the Earth that is polarized (on average)
can be treated as being from the jet axis to the direction
toward the observer, and so the only cosines of scattering angles
for this flux are $\mu$ and $-\mu$:  i.e., 
the cosines for the near and far jets
from the line of sight in the direction to the observer (see \S~2).
(For the reflection phase this approximation is justified
by our assumption of $\Delta\theta$ small. 
We justify it for the radial and optically-thin phases in \S~3.4.)
The approximation is, of course, somewhat gross.
The unpolarized (on average) scattered flux toward the observer we approximate 
as coming from an isotropic angular distribution, 
and so the cosine of the scattering angle is unneeded and unspecified. 

    Although we treat the electron scattering as coherent
in the jet frame, the supernova flux that scatters off the jet and
is emitted toward the observer at rest-frame wavelength
$\lambda$ is redshifted from the rest-frame wavelength
$\lambda'$ that it had on emission from the supernova photosphere.
We call this shifting the redshift effect.
It is caused by macroscopic-velocity-field Doppler shifts.
The approximate relation between the two wavelengths
from the non-relativistic Doppler formula is
\begin{equation}
\lambda=\lambda_{\pm}'[1+\beta_{\rm scat}(1\mp\mu)] \,\, , 
\end{equation}
where the upper and lower cases are for scattering off the near
and far jets, respectively, from the observer:  recall
$\mu$ is the cosine of the angle of axis of the jets measured 
from the line of sight
axis in the direction of the observer.
The $\beta_{\rm scat}$ velocity was defined in \S~2 (see eq.~[14])
as the
characteristic radial velocity of jet scattering:  for
the reflection phase $\beta_{\rm scat}\approx \beta_{\rm in}$
since the jet is very optically thick, and so 
$\tau_{\rm mean}/\tau_{\rm rad}<<1$.
There actually will be somewhat more redshift than given
by equation~(29) caused by multiple scattering in the
homologously expanding medium of jet.
In the reflection phase those scatterings are close together in
velocity space, and so probably cause negligible extra Doppler shift;
in later phases such shifts may be significant, but we will neglect
them for simplicity.
We actually will use the approximate inverse of equation~(29) in our formalism
below:
\begin{equation}
\lambda_{\pm}'=\lambda[1-\beta_{\rm scat}(1\mp\mu)] \,\, , 
\end{equation}
which is an inverse formula good to 1st order in $\beta_{\rm scat}$.

    To calculate the net jet-scattered flux and the net Stokes $Q$ parameter 
for this flux in the reflection phase we will, as mentioned at the beginning
of \S~2, assume the inner ends of the jets as hemispheres
with round sides toward the center of the supernova.
This assumption may be physically plausible and it
allows an analytical treatment.
But it does contradict our basic picture of the jet ends
as discussed at the beginning of \S~2.
 
    Let the cross sectional area of the hemispheres be $A$.
Toward the observer the reflecting face of the near jet hemisphere
will be crescent-shaped and that of the far jet hemisphere will be gibbous.
The faces of the hemispheres on sky have projected areas
\begin{equation}
             {1\over2}(1\mp\mu)A  \,\, , 
\end{equation}
where the upper and lower cases are
for the near and far jets, respectively.
Note that the summed projected area of the jets is just $A$.

    Approximating the supernova photosphere as a distant point
source, the hemispheres are wholly illuminated by the direct
supernova flux.
The direct supernova luminosity (emitted at wavelength
$\lambda_{\pm}'$) absorbed per unit projected
area (i.e., projected area as seen from the supernova center) by a jet is
\begin{equation}
{L_{\rm dir}(\lambda_{\pm}')\Omega_{\rm fr}(\Delta\theta)\over A} \,\, .
\end{equation}
An approximate expression for the reflected specific intensity emitted 
by a hemisphere in the direction toward the observer is
\begin{equation}
{L_{\rm dir}(\lambda_{\pm}')\Omega_{\rm fr}(\Delta\theta)\over 2\pi A}
  \left[g(\mu)q + (1-q)\right] \,\, .
\end{equation}
where the $g(\mu)q$ term accounts for the once-scattered, polarized photons,
and the $(1-q)$ for the multiply scattered, unpolarized photons.
Recall we assumed that we can approximate all cosines of scattering angles
for the once-scattered photons by the $\mu$ and $-\mu$.
We only need the $\mu$ in equation~(33) for both jets since 
the phase factor formula only depends on the square of 
the scattering angle cosine (see eq.~[28]).
The denominator $2\pi A$ in equation~(33) follows from the following
considerations.
If the energy absorbed by one jet from the direct flux from the bulk supernova 
were emitted by specific intensity beams that were constant with
angle of emission and surface location from a 
sphere of radius $r$, then the denominator would be $4\pi^{2}r^{2}$,
where the second $\pi$ factor accounts for the difference between flux
(using in the term in its formal radiative transfer sense) and
specific intensity.
In our case, all the absorbed direct flux energy gets radiated 
out of a hemisphere and not a sphere, and so
the numerator is set to $2\pi^{2}r^{2}$ or $2\pi A$ taking $r$ to be
the radius of the hemisphere.

    Equation~(33) is approximate because we have assumed a remote
point photosphere and because of the approximate nature of the $q$ factor.
It is also approximate because we assume that the energy collected from 
the bulk supernova is spread evenly over the hemispherical surface.
Actually the edges of the surface get less energy per surface
area than the center because of the hemispherical curvature:
the reflection specific intensity should decrease as one moves from
center to edges.
However, the overall reflected flux works out nearly correctly
as we show below probably because of the compensation we get when summing
up all the flux from both jet hemispheres emitted toward the observer.

    Now the flux at Earth from an object emitting
specific intensity $I$ is $f=\int I\,d\Omega=\int I\,dA/D^{2}$, 
where $d\Omega$ is the differential bit of solid angle at Earth that the object 
subtends and $dA$ is a differential bit of the
projected area of the object on the sky (e.g., \citeauthor{mihalas1978}, p.~11).
Given this fact and making use of equations~(17), (30), (31), 
and~(33), it follows that the total 
jet flux measured at Earth at the rest-frame wavelength
$\lambda$ for the reflection phase is
\begin{eqnarray}
f_{\rm reflection}(\lambda)
&=&\left[ f_{\rm dir}(\lambda_{+}')(1-\mu)
         +f_{\rm dir}(\lambda_{-}')(1+\mu)
         \right]\Omega_{\rm fr}(\Delta\theta) \nonumber    \\
&&\qquad\qquad
   \left[ g(\mu)q+(1-q)\right] \,\, . 
\end{eqnarray}
We note that if $\lambda_{+}'$ and $\lambda_{-}'$ are both set
to $\lambda$ and $q$ to zero, then
\begin{eqnarray}
f_{\rm reflection}(\lambda)
   =2{L_{\rm dir}(\lambda)\Omega_{\rm fr}(\Delta\theta)
                          \over 4\pi D^{2} }  \,\, . 
\end{eqnarray}
This last result is the flux at the Earth if all the luminosity captured
by the jets were radiated isotropically.
That we get this result verifies that we are accounting for the energy
approximately correctly in equation~(33) for the reflected specific
intensity.

     From equation~(34) and making use of our understanding of equation~(28) 
for the Rayleigh phase function, we find that the net Stokes $Q$ parameter 
for the reflected flux is
\begin{eqnarray}
Q_{\rm reflection}(\lambda)
&=&\left[ f_{\rm dir}(\lambda_{+}')(1-\mu)
         +f_{\rm dir}(\lambda_{-}')(1+\mu)
         \right]\Omega_{\rm fr}(\Delta\theta)  \nonumber \\
&&\qquad\qquad
     \left({3\over4}\right)\left(\mu^{2}-1\right)q
      \xi(\Delta\theta)   \,\, , 
\end{eqnarray}
where we have introduced the correction factor
$\xi(\Delta\theta)$ (see eq.~[26]) that accounts for 
effect of the range of scattering planes
on the Stokes $Q$ parameter.
Note the $(1-q)$ term in equation~(34) accounts for unpolarized flux, 
and thus the $x$- and $y$-fluxes it governs are equal and it cancels out of
$Q_{\rm reflection}$.

\subsection{Radial and Optically Thin Phases}
  
    In the ideal picture of the radial phase,
a jet is optically thick in the radial direction, and so
no photons can escape out the outer end of the jet.
On the other hand, we assume the jet is optically
thin perpendicular to its axis.
With this setup,  
radially-outward scattered photons remain
trapped in the jet as do the radially-inward scattered ones (except
at inner end of the jet) and off-radially scattered
photons are assumed to escape without scattering again.
Recall that forward and backward scattering are unpolarizing
as determined by the Rayleigh phase matrix (see \S~3.3),
and thus the radially scattered photons remain unpolarized.
As mentioned in \S~3.3, we assume that all scattering toward the Earth can
be treated as happening along the jet axis, and so the
cosines of scattering angles are again $\mu$ and $-\mu$.
Thus there is no averaging over a range of
scattering cosines.
We again will include our correction factor $\xi(\Delta\theta)$
(see \S~3.3, eq.~[26])
to account for the range of scattering planes.
Recall the scattering plane defined by the jet axis and the line of sight
to the observer is the fiducial plane.

    Given aforesaid picture, all the photons that
escape the jet are distributed in angle
and polarized as if they had undergone a single electron scattering
event (i.e., their last scattering which was off-radial
or radially-inward)---except that in our picture there
is no flux scattered radially-outward and perhaps
less scattered radially-inward since some photons
should random walk far out in the radial direction.
The exception tends to cause more polarized emission than
predicted by the Rayleigh phase matrix for a single scattering.
Trying to invent a correction factor for this effect that is of the right size
seems futile given the crudity of the model.

     To handle the phase where the jet becomes optically
thin radially (i.e., the optically-thin phase), we include the factor
$1-\exp(-\tau_{\rm rad})$ in our expressions for flux and Stokes $Q$
parameter:   $1-\exp(-\tau_{\rm rad})$ is the fraction of photons
that scatter in the jet at least once:  it goes to
zero as $\tau_{\rm rad}$ goes to zero.
Thus the expressions we develop in this section
will handle both the radial phase and optically-thin phase.

     The ideal picture for radial phase is very polarizing
for the direct supernova flux that passes into the jet because
of the off-radial direction is optically thin.
To make the expressions a bit more realistic we will account for
finite optical depth perpendicular to the axis in the following
way.
As defined by equation~(14) in \S~2, $\beta_{\rm scat}$ is the mean
velocity position that photons incident on a jet reach
for their first scattering.
The perpendicular-to-axis optical depth from axis to
jet surface at $\beta_{\rm scat}$ is 
$\tau_{\rm off}$ which is approximately given by equation~(15) in \S~2.
We take $\exp(-\tau_{\rm off})$ as the characteristic
probability that an off-radially scattered photon
escapes without scattering again.
We take $1-\exp(-\tau_{\rm off})$ as the probability
that an off-radially scattered photon scatters again.
We assume that photons that scatter multiple times
in non-radial directions have 
on average an isotropic angular distribution and
zero polarization:  the multiple off-radial scattering
should average away anisotropy and net polarization.

      Now recall that the direct supernova luminosity 
(emitted at wavelength $\lambda_{\pm}'$) passing into a jet is 
$L_{\rm dir}(\lambda_{\pm}')\Omega(\Delta\theta)$.
Multiplying this by $1-\exp(-\tau_{\rm rad})$ gives the
luminosity that interacts in the jet:
i.e., 
$L_{\rm dir}(\lambda_{\pm})\Omega(\Delta\theta)[1-\exp(-\tau_{\rm rad})]$.
We now assume that all this interacting luminosity escapes 
with a normalized angular distribution
determined by 
\begin{equation}
{g(\mu_{\rm gen})
\exp(-\tau_{\rm off})+[1-\exp(-\tau_{\rm off})] \over 4\pi} \,\, : 
\end{equation}
i.e., a combination of Rayleigh and isotropic scattering with the
relative amount of each determined by the probability of no second
off-radial
scattering $\exp(-\tau_{\rm off})$.
Thus the luminosity per solid angle in the direction of the Earth from
a jet at wavelength $\lambda$ is 
\begin{equation}
L_{\rm dir}(\lambda_{\pm}')\Omega(\Delta\theta)[1-\exp(-\tau_{\rm rad})]
{1\over 4\pi}
\left\{g(\mu)\exp(-\tau_{\rm off})+[1-\exp(-\tau_{\rm off})]\right\}
\,\, , 
\end{equation}
where again the upper case of $\lambda_{\pm}'$ is for the near jet and the
lower case for the far jet, $\lambda_{\pm}'$ is determined by equation~(30)
(see \S~3.3), and $g(\mu)$ applies to both
near and far jets since this formula depends only 
on the square of the scattering angle cosine (see eq.~[28] in \S~3.3).
If we multiply the luminosity per unit solid angle by
$d\Omega_{\rm Earth}/(D^{2}d\Omega_{\rm Earth})$, where
$d\Omega_{\rm Earth}$ is a differential bit of solid angle at a jet
subtended by the Earth and $D^{2}d\Omega_{\rm Earth}$ is the
projected area of the Earth on the sky as seen from the jet, 
we get the flux at Earth from one jet:  i.e.,
\begin{eqnarray} 
&&{L_{\rm dir}(\lambda_{\pm}')\Omega(\Delta\theta)\over 4\pi D^{2} }
[1-\exp(-\tau_{\rm rad})]
\left\{g(\mu)\exp(-\tau_{\rm off})+[1-\exp(-\tau_{\rm off})]\right\}
  \nonumber \\
&=& f_{\rm dir}(\lambda_{\pm}')\Omega(\Delta\theta)
[1-\exp(-\tau_{\rm rad})]
\left\{g(\mu)\exp(-\tau_{\rm off})+[1-\exp(-\tau_{\rm off})]\right\} \,\, , 
\end{eqnarray}
where we have used equation~(17) from \S~3.1.
The total flux jet flux at Earth at wavelength $\lambda$ for the
radial and optically-thin phases
\begin{eqnarray}
f_{\rm radial/optically\hbox{-}thin}(\lambda)
&=&\left[ f_{\rm dir}(\lambda_{+}')
         +f_{\rm dir}(\lambda_{-}')\right] \Omega_{\rm fr}(\Delta\theta) 
	   \left[1-\exp(-\tau_{\rm rad})\right] \nonumber \\
&&\quad
   \biggl\{g(\mu)\exp(-\tau_{\rm off})
         +\left[1-\exp(-\tau_{\rm off})\right]\biggr\}  \,\, .
\end{eqnarray}
Note that the $1\mp\mu$ factors multiplying
the direct fluxes that occur in the reflection phase
expressions (see eq.~[34] and~[36] in \S~3.3) do not appear in
equations~(39) and~(40):
in the ideal radial picture all parts of the jets send
light to the observer and not just the Earth-side parts of the inner ends.
The ideal radial phase is when $\tau_{\rm off}\to0$ or
$\exp(-\tau_{\rm off})\to1$:  recall, the $\exp(-\tau_{\rm off})$ and
$1-\exp(-\tau_{\rm off})$ factors correct for the less-than-ideal
radial picture situation. 

    In developing the equation~(40), we are assumed the jets are 
not too closely aligned with the line of sight
which would be at least approaching the occultation situation:  
as discussed in \S~2,
we are not concerning ourselves with the occultation situation
in this paper.

     From equation~(40) and again making use of our understanding of equation~(28)
for the Rayleigh phase function (see \S~3.3), 
we find that the net Stokes $Q$ parameter for the observer-scattered flux is
\begin{eqnarray}
Q_{\rm radial/optically\hbox{-}thin}(\lambda)
&=&\left[ f_{\rm dir}(\lambda_{+}')
         +f_{\rm dir}(\lambda_{-}')
        \right]
\Omega_{\rm fr}(\Delta\theta)\left[1-\exp(-\tau_{\rm rad})\right] 
   \nonumber \\
&&\qquad \left({3\over4}\right)\left(\mu^{2}-1\right)\exp(-\tau_{\rm off})  
         \xi(\Delta\theta) 
    \,\, ,  
\end{eqnarray}
where we have again introduced the correction factor
$\xi(\Delta\theta)$ (see eq.~[26] in \S~3.3) to account for
the range of scattering planes.
Note the $1-\exp(-\tau_{\rm off})$ term in equation~(40) 
accounts for unpolarized flux, and thus the $x$- and $y$-fluxes 
it governs are equal 
and it cancels out of $Q_{\rm radial/optically\hbox{-}thin}$.

\subsection{General Expressions for All Phases}

     The flux and Stokes $Q$ parameter expressions of \S\S~3.3 and~3.4
were developed from idealized limiting-case pictures.
The state of the jets in the times between when these pictures
apply is complex, but it would be useful to have general
expressions that cover the intermediate times as well.
We will construct general expressions that are valid in
the reflection and ideal radial/optically-thin phases and act
as  interpolation formulae for the intermediate times.
(The ideal radial phase is when $\tau_{\rm off}\to0$ recall.)
We are guided somewhat by an intermediate phase picture.

    The reflection phase picture (see \S~3.3) predicts that the
fraction $q$ of the photons incident on a jet undergo 
a single scattering and are highly polarized.
The rest of the photons (i.e., the fraction $1-q$) are multiply scattered and emerge
unpolarized (or rather randomly polarized and so yielding no net
polarization).
The radial phase ideally assumes that all scattered photons are
scattered as if they had undergone a single scattering
event and is therefore a more polarizing phase ideally.
Less ideally, we estimate the polarized fraction of the escaping
scattered photons
is $\exp(-\tau_{\rm off})$, where $\tau_{\rm off}$ is the
off-radial optical depth (i.e., the optical depth from the
axis to the surface along a perpendicular to the axis) 
at the mean radial optical for the first scattering of a radial photon
(i.e., at $\beta_{\rm scat}$).
We will argue that there can be no local minimum
in polarization between the reflection and radial phases.
First, it is plausible that the ratio of singly scattered
to all scattered photons should be roughly constant as the
the reflection phase changes to the radial phase:  both
phases are optically thick in the radial direction and
all photons are scattered:  the photons just tend to penetrate
more deeply as the jets expand with time. 
Just as the jet becomes optically thin perpendicular to
the axis (i.e., $\tau_{\rm off}\to0$)
the fraction of multiply scattered photons that are polarized
(i.e., polarized in a non-random way)
should increase from zero to all multiply scattered photons ideally:
recall our description of the ideal radial phase scattering at the 
beginning of \S~3.4.
Now net polarization is set by the non-randomly polarized, scattered photons.
Since these should increase in abundance as we go from the reflection
phase to the ideal radial phase, we should expect polarization
to increase monotonically. 

    We now can derive general expressions that handle the
the reflection and radial/optically-thin phases and interpolate
between them. 
The discussion just above suggests replacing the $q$ and
$1-q$ factors in the reflection phase equations for flux and Stokes
$Q$ parameter  (see eq.~[34] and~[36] in \S~3.3) 
and
the $\exp(-\tau_{\rm off})$ and $[1-\exp(-\tau_{\rm off})]$ factors
in the counterpart radial/optically-thin phase expressions
(see eq.~[40] and~[41] in \S~3.4) 
by, respectively,
\begin{eqnarray}
w&=&q+(1-q)\exp(-\tau_{\rm off})
   =\cases{q \,\, & for $\tau_{\rm off}\to\infty$;\cr 
           1 \,\, & for $\tau_{\rm off}\to 0$ \cr}  \\ 
\noalign{\noindent and}
1-w&=&(1-q)[1-\exp(-\tau_{\rm off})]
   =\cases{1-q \,\, & for $\tau_{\rm off}\to\infty$;\cr
           0 \,\,   & for $\tau_{\rm off}\to 0$,  \cr}  
\end{eqnarray}
since the change from reflection to radial phases can plausibly
be identified as when $\tau_{\rm off}$ drops below about 1.
With these replacements the reflection phase 
and radial/optically-thin phase expressions become more alike.
If we add a factor $[1-\exp(-\tau_{\rm rad})]$ to the
reflection phase equations for flux and Stokes
$Q$ parameter, the expressions for the
reflection and radial/optically-thin phases become identical, except for the
$1\mp\mu$ factors in the reflection phase expressions. 
To complete the general expressions, we must find away to
smoothly change the $1\mp\mu$ factors into $1$'s as the reflection phase
changes to the radial/optically-thin phase.
 
    The $(1\mp\mu)$ factors were based on the very
special, but heuristically useful, assumption that
the inner ends of the jets could be treated as hemispheres
with round sides toward the supernova center.
We do not put much weight on this hemisphere assumption
and have no strong guidance on how to effect
the change from $1\mp\mu$ factors to $1$'s
as the reflection phase changes to the radial phase.
Therefore, no elaborate procedure is warranted. 
One possibility is multiply the $\mu$'s in the $1\mp\mu$ factors by
\begin{equation}
w_{\mu}=\cases{\displaystyle
  1-{1\over\tau_{\rm off}}\,\,   &for $\tau_{\rm off}>1$; \cr
\noalign{\smallskip}
  0\,\,  &for $\tau_{\rm off}\leq 1$ \cr}
\end{equation}
so that one has $1\mp w_{\mu}\mu$ instead of $1\mp\mu$ in the expression.
The $w_{\mu}$ factor will give smooth transition between the phases
and at what can be regarded as the characteristic 
end of the reflection phase when $\tau_{\rm off}=1$ 
the $w_{\mu}$ factor becomes a 0 and the
$1\mp\mu$ factors have effectively changed to exactly $1$'s.

    One could, of course, devise alternative definitions of
$w_{\mu}$ that change from 1 to 0 as the phases change.
A very simple alternative is just to set $w_{\mu}=0$ at all times.
This simple alternative is warranted because we do not
take the hemisphere assumption too seriously and also because
the exact value $w_{\mu}\mu$ is not too important if 
$f_{\rm dir}(\lambda_{+}')$ and $f_{\rm dir}(\lambda_{-}')$
do not differ too much which may often be the case.
Also in actual cases where polarization is high, $\mu$ is likely 
to be closer to 0 than to 1 which would tend to make the
exact value of $w_{\mu}$ of little importance.

    Below we will insert $w_{\mu}$ factors to account for the
transition of $1\mp\mu$ factors to $1$'s, but we make no
final specification of $w_{\mu}$ factors.
In all our demonstration calculations in \S~4, we take $\theta=90^{\circ}$
(i.e., $\mu=0$) and the value of $w_{\mu}$ factors is irrelevant. 

     Making all the changes to the expressions for the
reflection and radial/optically-thin phases that we suggested above
and replacing $1\mp\mu$ by $1\mp w_{\mu}\mu$ in the
reflection phase expressions and inserting $1\mp w_{\mu}\mu$ in
the radial/optically-thin phases appropriately, makes
the reflection and radial/optically-thin phase expressions identical
and these identical expressions are our general expressions
for the jet-scattered flux and its Stokes $Q$ parameter.
Explicitly, the general expressions are
\begin{eqnarray}
f_{\rm jet}(\lambda)
&=&\left[ f_{\rm dir}(\lambda_{+}')(1-w_{\mu}\mu)
         +f_{\rm dir}(\lambda_{-}')(1+w_{\mu}\mu)
         \right]  \nonumber \\
&&\qquad
\Omega_{\rm fr}(\Delta\theta)\left[1-\exp(-\tau_{\rm rad})\right]
 \left[ g(\mu)w+(1-w)\right]
\end{eqnarray}
and
\begin{eqnarray}
Q_{\rm jet}(\lambda)
&=&\left[ f_{\rm dir}(\lambda_{+}')(1-w_{\mu}\mu)
         +f_{\rm dir}(\lambda_{-}')(1+w_{\mu}\mu)
         \right]     \nonumber \\
&&\qquad
\Omega_{\rm fr}(\Delta\theta)\left[1-\exp(-\tau_{\rm rad})\right]
   \left({3\over4}\right)\left(\mu^{2}-1\right)w
    \xi(\Delta\theta) \,\, . 
\end{eqnarray}
These expressions interpolate smoothly between the reflection phase
and the ideal radial phase (where $\tau_{\rm off}\to0$) and
also handle the optically-thin phase, of course.
This can be seen by comparing in detail the expressions to equations~(34) 
and~(36) in \S~3.3 and equations~(40) and~(41) in \S~3.4.

     For actual analysis we would like the ratio $R$
of jet-scattered to direct flux (which we call the relative
scattered flux) and the polarization expression for the total flux.
Dividing equation~(45) by 
$f_{\rm dir}(\lambda)$ gives the relative scattered flux
\begin{eqnarray}
R(\lambda)
&=&{f_{\rm jet}(\lambda)\over f_{\rm dir}(\lambda)}
 =\left[{ f_{\rm dir}(\lambda_{+}')(1-w_{\mu}\mu)
          +f_{\rm dir}(\lambda_{-}')(1+w_{\mu}\mu)
	 \over f_{\rm dir}(\lambda) } \right]  \nonumber   \\
   &&\qquad\qquad\qquad\qquad
\Omega_{\rm fr}(\Delta\theta)\left[1-\exp(-\tau_{\rm rad})\right]
    \left[ g(\mu)w+(1-w)\right]   \,\, .
\end{eqnarray}
For polarization we divide the absolute value of the $Q_{\rm jet}(\lambda)$ 
expression
by total flux $f_{\rm dir}(\lambda)[1+R(\lambda)]$ 
and multiply by 100$\,$\% since
polarization is conventionally given as a percentage:
\begin{eqnarray}
P(\lambda)
&=&\left[{ f_{\rm dir}(\lambda_{+}')(1-w_{\mu}\mu)
          +f_{\rm dir}(\lambda_{-}')(1+w_{\mu}\mu)
	  \over f_{\rm dir}(\lambda)[1+R(\lambda)]} \right]     \nonumber \\
  &&\qquad\qquad
\Omega_{\rm fr}(\Delta\theta)\left[1-\exp(-\tau_{\rm rad})\right]
   \left({3\over4}\right)\left(1-\mu^{2}\right)w\xi(\Delta\theta) 
    \times 100\,\%  \,\, . 
\end{eqnarray}
Recall we assume that the direct supernova flux is unpolarized and
therefore has zero Stokes $Q$ parameter:  thus equation~(48) gives
the total supernova polarization for the bipolar jet model.
The $1/[1+R(\lambda)]$ factor in the polarization expression is 
probably usually negligibly different from 1 for cases where our
approximations are best:  i.e., when jets subtend relatively
little solid angle (i.e., when $\Omega_{\rm fr}(\Delta\theta)$ is
small), and thus are relatively narrow and 
scatter relatively little flux.

   If one is trying to obtain $R$ and $P$ from observed flux,
then one can substitute 
$f_{\rm obs}(\lambda)/[1+R(\lambda)]$,
$f_{\rm obs}(\lambda_{+}')/[1+R(\lambda_{+}')]$, and
$f_{\rm obs}(\lambda_{-}')/[1+R(\lambda_{-}')]$ into
equations~(47) and~(48) 
for $f_{\rm dir}(\lambda)$, $f_{\rm dir}(\lambda_{+}')$, and
$f_{\rm dir}(\lambda_{-}')$, respectively.
Usually one can then drop the inserted $1/(1+R)$ factors as being negligible
and then no complex solution for $R$ as a function of wavelength is required.
In fact, the inserted $1/(1+R)$ factors probably often partially cancel out 
of the equations: 
they exactly cancel out if $R$ is constant with wavelength.
A case where $R$ might become large and where a complex solution might
be needed for high accuracy is in the trough features of
P~Cygni lines where $f_{\rm dir}(\lambda)$ becomes small
while $[f_{\rm dir}(\lambda_{+}')+f_{\rm dir}(\lambda_{-}')]$ could stay
large due to the redshift effect. 
But given that the bipolar jet model is crude, complex solutions for
$R$ are probably a waste of effort almost always. 

     Our general expressions for $R$ and
polarization $P$ depend on 5~independent basic parameters:
$\theta$ (the angle of the jet axis from the
line of sight) or $\mu=\cos\theta$, 
$\Delta\theta$ (the half-opening angle of
the jets), $\beta_{\rm in}$ (the inner end of the
jet in velocity), $b=\Delta\beta_{\rm rad}/\beta_{\rm in}$
(the ratio of jet velocity length to inner end velocity of jet),
and $\tau_{\rm rad}$ (or if the time-delay effect is included
$t_{\rm red}$:  see \S~3.2).
The $\tau_{\rm rad}$ parameter can be resolved into time $t$ and the
$\tau_{\rm rad,0}t_{0}^{2}$ parameter 
(see eq.~[9] in \S~2) for an analysis of the
time evolution of polarization:  in this case $\tau_{\rm rad,0}t_{0}^{2}$
replaces $\tau_{\rm rad}$ as one of the 5 independent basic parameters.
The $q$ parameter, which is an argument of the $w$ factor
(see eq.~[42] and~[43] above) is also independent,
but we consider its value as fixed at $1/4$ (see \S~3.3, esp.~eq.~[23]), 
and so exclude it from our list of independent basic parameters.
The non-basic parameters in the $R$ and $P$ depend on the basic parameters
as follows:  
$\lambda_{\pm}'$ depends on $\theta$, 
$\beta_{\rm in}$, $b$, and $\tau_{\rm rad}$ 
(see eq.~[13] and [14] in \S~2 and eq.~[30] in \S~3.3);
$w$ depends on $\Delta\theta$, $b$, and $\tau_{\rm rad}$
(and on $q$ too)
(see eq.~[13] and~[15] in \S~2 and eq.~[42] and~[43] above);
$w_{\rm\mu}$ using our first prescription
depends on $\Delta\theta$, $b$, and $\tau_{\rm rad}$,
(see eq.~[13] and~[15] in \S~2 and eq.~[44] above) and
on nothing if we set it to zero.

    Given the 5 independent basic parameters,
one can correct the synthetic supernova spectrum $f_{\rm dir}(\lambda)$ of
any spherically symmetric
calculation to include the additional flux from a jet by
multiplying this spectrum by $1+R$, where $R$ itself is evaluated using
the synthetic $f_{\rm dir}(\lambda)$.
An observed spectrum (which naturally includes jet flux if there
are jets) can by corrected to the direct supernova spectrum
by dividing by $1+R$ or multiplying by $1-R$ assuming $R$ small.
In this case, $R$ is evaluated using the observed $f_{\rm obs}(\lambda)$
(as discussed above)
which, of course, includes jet flux if there is any:  this,
however, gives a direct flux that is first order correct in small $R$.

   There is considerable degeneracy among the 5 independent basic parameters
and finding a uniquely good set for a single continuum polarization
observation would be difficult.
A fit to an observed time evolution of continuum polarization
may break some of the degeneracy.

    For example, consider $\theta$ and $\Delta\theta$.
The polarization increases with $\theta$ up to $\theta$'s 
maximum value
of $90^{\circ}$ degrees through the $1-\mu^{2}$ factor.
But the polarization also scales with 
$\Omega_{\rm fr}(\Delta\theta)\xi(\Delta\theta)$ which is 
the most obvious manifestation of the $\Delta\theta$ parameter:  
$\Omega_{\rm fr}(\Delta\theta)\xi(\Delta\theta)$
increases with $\Delta\theta$ from zero 
for $\Delta\theta=0^{\circ}$ to a maximum of $0.1046$ at 
$\Delta\theta\approx 56.48^{\circ}$
(which is beyond where our picture of a narrow jet has plausible
validity) and then decreases with $\Delta\theta$ to zero at 
$\Delta\theta=90^{\circ}$.
Clearly, $\theta$ and $\Delta\theta$ could be varied in
a compensating way while holding the polarization constant or nearly.
Thus, there is a strong time-independent degeneracy between $\theta$ and
$\Delta\theta$.

     Now $\Delta\theta$ also affects the $w$ factor 
(see eq.~[42] above) through the $\tau_{\rm off}$ optical
depth (see eq.~[15] in \S~2).
Since $w$ varies in time due to the time variation of $\tau_{\rm rad}$
(see eq.~[13] and [15] in \S~2), the degeneracy of $\Delta\theta$ and
$\theta$ may be broken by fitting to an observed time evolution of
continuum polarization.
However, two other parameters come into $\tau_{\rm off}$:
$b$ (see eq.~[15] in \S~2)
and $\tau_{\rm rad,0}t_{\rm,0}^{2}$ (which replaces $\tau_{\rm rad}$
as an independent basic parameter in a time-dependent calculation)
(see eq.~[9] in \S~2), and these can be used to 
compensate variations in $\Delta\theta$ to some degree.
The situation is further complicated by the fact that the time
variation of $\tau_{\rm rad}$ also directly affects the polarization through
the $1-\exp(-\tau_{\rm rad})$ factor.

    Clearly, even when studying an observed time evolution of  
continuum polarization, parameter degeneracy will be a problem in finding 
a uniquely good set of the 5 independent basic parameters.
Any constraints on the jet structure from non-polarimetric observations
or theory would help to break the degeneracy.

    Another parameter degeneracy example involves the redshift effect.
The redshift effect depends on $\theta$ (through $\mu$)
and $\beta_{\rm scat}$ (see eq.~[30] in \S~3.3).
Now $\beta_{\rm scat}$ depends on $b$, $\beta_{\rm in}$,
and $\tau_{\rm rad}$ (see eq.~[13] and [14] in \S~2).
The redshift effect manifests itself most prominently in polarization 
spectra:  we study this manifestation in \S~4.3 and~4.4.
If $\theta$, $b$, and $\tau_{\rm rad}$ can be determined from the time evolution
of the continuum polarization then that breaks the degeneracy
with $\beta_{\rm in}$ which can
be determined from analysis of polarization spectra 
since $\beta_{\rm in}$ has a strong effect on the redshift
effect (see eq.~[14] in \S~2 and eq.~[30] in \S~3.3).

    Even if a unique set of the 5 independent basic parameters 
cannot be obtained from
polarimetric observations, it would still be very interesting
if the bipolar jet model could be shown to be adequate for
those observations.

\section{Demonstration Calculations}

    In this section, we present demonstration calculations for
continuum polarization and polarization spectra.
In the continuum polarization case, it is of especial interest to see
if the bipolar jet model can yield a continuum polarization time evolution
that rises to a maximum and then declines with time (see \S~4.2)
and if it can yield the inverted P~Cygni polarization profiles of lines
(see \S\S~4.3 and~4.4).

    For simplicity we neglect the time-delay effect (see \S~3.2) in
all our calculations.
This means that we use a single reduced time $t_{\rm red}$ (which
specifies the radial optical depth $\tau_{\rm rad}$ through
eq.~[10] in \S~2) to calculate the polarization for each epoch or
we use $\tau_{\rm rad}$ as one of the independent basic parameters.
Also in all our calculations, we set $\theta=90^{\circ}$
(i.e., the jets are aligned perpendicular to the line of sight):  recall
$\theta$ is one of the 5 independent basic parameters (see \S~3.5). 
This $\theta$ is the maximally polarizing choice as one can
see from equation~(48) in \S~3.5, where the main $\theta$ dependence
is through $1-\mu^{2}$ where $\mu=\cos\theta$. 
(The $\theta$ parameter also affects the polarization through the
redshift effect: see equation~[30] in \S~3.3.)

    In \S~4.1 we consider continuum polarization with 
varying $\Delta\theta$ and in \S~4.2 continuum polarization as
a function of time.
In \S~4.3, we discuss polarization line and continuum features.
Synthetic polarization spectrum calculations are presented in
\S~4.4. 

\subsection{Continuum Polarization Calculations with Varying 
            $\Delta\theta$}

    For the strictly continuum polarization calculations in this
subsection and \S~4.2, we assume
the direct flux from the bulk supernova is constant with
wavelength:  this is the neutral choice.
Thus we have 
$f_{\rm dir}(\lambda_{+}')=f_{\rm dir}(\lambda_{-}')
 =f_{\rm dir}(\lambda)$ for these calculations 
and in equation~(48) in \S~3.5 the factor 
\begin{equation}
{ f_{\rm dir}(\lambda_{+}')(1-\omega_{\mu}\mu)
 +f_{\rm dir}(\lambda_{-}')(1+\omega_{\mu}\mu)
\over f_{\rm dir}(\lambda)}=2  \,\, : 
\end{equation}
note that the values of $\omega_{\mu}$ and $\mu$ are 
irrelevant in the flux ratio factor for a constant continuum flux.
The assumption of wavelength-independent direct flux 
effectively eliminates the redshift
effect from the calculations (see eq.~[30] in \S~3.3).
This means that the independent basic parameter $\beta_{\rm in}$
and the dependent parameter $\beta_{\rm scat}$ (see eq.~[14]
in \S~2) never come into the calculations.
Thus, we do not specify values of $\beta_{\rm in}$ in this section or
in \S~4.2.

     In this subsection, we are interested in seeing 
how continuum polarization varies
with $\Delta\theta$ with other parameters chosen for a high 
polarization case in the radial phase.
To get a large polarization we set $b=\infty$ 
(which makes the jets infinitely long compared to distance from the supernova center
to inner jet surface:  see eq.~[4] in \S~2) 
and $\tau_{\rm rad}>>1$ (which means all photons
entering the jet are scattered).
These choices make $\tau_{\rm rad}/b=0$, $\tau_{\rm mean}=1$,
and $\tau_{\rm off}=\Delta\theta$ (see eq.~[13] and~[15] in \S~2)
and $1-\exp(-\tau_{\rm rad})\approx 1$:
a fairly small $\tau_{\rm off}$ and $1-\exp(-\tau_{\rm rad})$
nearly 1 produce high polarization (see eq.~[48] and~[42] in \S~3.3). 
(As we will show in Figure~1 in \S~4.2, slightly, but distinctly, 
higher continuum polarizations than for $\tau_{\rm rad}>>1$
can be achieved for $b=\infty$ and $\Delta\theta\gtrsim30^{\circ}$ 
as $\tau_{\rm rad}$ declines with time toward the neighborhood of 1.)

     For the specified 3 independent basic parameters 
(e.g., $\theta=90^{\circ}$,
$b=\infty$, and $\tau_{\rm rad}>>1$) and independent
basic parameter $\beta_{\rm in}$ unspecified since it is irrelevant,
Table~\ref{table-1} shows, as a function of 
the independent basic parameter
$\Delta\theta$, the runs of $\Omega_{\rm fr}(\Delta\theta)$, 
$\xi(\Delta\theta)$, $w$,
$R$, and polarization $P$.
Equation~(48) in \S~3.5 shows how polarization depends on
the dependent parameters
$\Omega_{\rm fr}(\Delta\theta)$, $\xi(\Delta\theta)$, $w$, and $R$ in
a direct sense.
The competing effects of $\Delta\theta$ through the
dependent parameters yield a polarization maximum of about $6.63\,$\% at
$\Delta\theta\approx 45.6^{\circ}$.
This opening angle is beyond where our picture of a narrow jet
has plausibility.
However a continuum polarization of $4\,$\% is reached at
$\Delta\theta\approx 24^{\circ}$ which is plausibly still a sufficiently narrow
jet to channel non-escaping photons mostly radially as 
in our ideal picture of the radial phase discussed in \S~3.4.

     The polarization result of $4\,$\% just mentioned
is encouraging for the bipolar jet model if it turns out that
polarization of this size (which is possibly the current
record continuum polarization for a supernova:
see \S~1.1) is required from the model to fit observations.
However, the result is obtained with high-polarizing choices for $\theta$,
$b$, and $\tau_{\rm rad}$.
For example if we make $b=0$, the jets become
geometrically short along their axis:  the radial channeling
effect of non-escaping scattered photons disappears.
The jets in this case remain in the reflection phase
with $\tau_{\rm off}=\infty$ and $w=q=1/4$
(see eq.~[15] in \S~2 and eq.~[42] in \S~3.5 and eq.~[23] in \S~3.3)
until they become optically thin radially.
(Having $w=1/4$ when the jets becomes radially
optically thin is an unphysical limiting case for the equation~[48]
since most photons that scatter once will escape without being scattered
again and therefore will not be depolarized by multiple scattering 
as $w=1/4$ implies.) 
Much lower polarizations than in the $b=\infty$ case would be found
since every explicit factor in equation~(48) is the same as for
the $b=0$ case, except 
that $w$ is distinctly smaller (i.e., $w=1/4$ as compared to 
the $w$ values in Table~\ref{table-1}) which distinctly decreases polarization
and that $R$ increases a bit which is because
of smaller $w$ and slightly decreases polarization.
With fiducial parameters used in Table~\ref{table-1}, except
that $b$ is chosen to be 0, the continuum polarization reaches a maximum of only
about $2.8\,$\% for $\Delta\theta=51^{\circ}$.  
For $\Delta\theta=24^{\circ}$, the polarization is only $1.3\,$\%.
     
     The effect of the $b$ parameter on the time evolution
of the polarization is significant. 
As long as $\Delta\theta\tau_{\rm rad}/b$ (with $\Delta\theta$
in radians, of course) is large, the off-radial escape
of highly polarized photons, except for reflected photons the inner end
of the jet, is small.
Recall this escape is governed in the model
by $\exp(-\tau_{\rm off})$ (see eq.~[42] and~[43] in \S~3.5), where
$\tau_{\rm off}=\Delta\theta(\tau_{\rm rad}/b+\tau_{\rm mean})$ 
(see eq.~[15] in \S~2).
Only when $\tau_{\rm off}$ becomes less than of order 1  
while $\tau_{\rm rad}$ is still significantly larger than 1 
does the ideal radial phase high polarizing effect 
that was discussed in \S~3.4 begin to be approached.
However, if $\tau_{\rm rad}$ is falling below $1$ when
$\tau_{\rm off}$ becomes of order 1, then polarization could
start falling because a decreasing number of photons are 
scattered at all in the jet.
Thus, we expect localized polarization maxima in time evolution
when $\tau_{\rm off}\lesssim 1$ (requiring
$\Delta\theta\tau_{\rm rad}/b\lesssim 1$), but $\tau_{\rm rad}\gtrsim 1$.
Our calculations bear this out:  see \S~4.2, esp.~Fig.~2.

\subsection{Calculations of Continuum Polarization as a Function of Time} 

    In Figures~1, 2, and~3, we show the time evolution of
the continuum polarization for representative $b$ values and
$\Delta\theta$ values with $\theta=90^{\circ}$.
Since we are assuming a wavelength-independent direct supernova
flux as in \S~4.1, 
$[{f_{\rm dir}(\lambda_{+}')(1-w_{\mu}\mu)
  +f_{\rm dir}(\lambda_{-}')(1+w_{\mu}\mu)]
  /f_{\rm dir}(\lambda)}=2$ and
$\beta_{\rm in}$ is unspecified since it is irrelevant
for a wavelength-independent direct supernova flux.
The time used is the reduced time $t_{\rm red}=1\sqrt{\tau_{\rm rad}}$
(see \S~2, esp.~eq.~[10]) which eliminates the need to specify
a $\tau_{\rm rad,0}t_{0}^{2}$ parameter (see eq.~[9] in \S~2).

    Figure~1 shows the $b=0$ cases (solid lines) 
and $b=\infty$ cases (dashed lines) for a range
of $\Delta\theta$ values including $46^{\circ}$ which yields
near maximum polarization for the $b=\infty$ case
(see Table~\ref{table-1}).
The only difference between the $b=0$ and $b=\infty$ cases is that
in the former $w=q=1/4$ for all times and in the latter
$w=q+(1-q)\exp(-\Delta\theta\tau_{\rm mean})$
(see eq.~[13] and~[15] in \S~2, eq.~[23] in \S~3.3, 
and eq.~[42] in \S~3.5).
In the $b=\infty$ cases, $w$ will rise monotonically
from $w=q+(1-q)\exp(-\Delta\theta)$ at early times
(when $\tau_{\rm rad}>>1$ and $\tau_{\rm mean}=1$:  see 
eq.~[13] and~[15] in \S~2)
to 1 at late times (when $\tau_{\rm mean}\to0$):  
$w$ in the $b=\infty$ cases will always be greater than $q=1/4$.

    The early time behavior of the $b=0$ cases is the first kind of
polarization plateau phase discussed at the end of \S~2.
The early time behavior of the $b=\infty$ cases is the second kind
of polarization plateau phase likewise discussed at the end of \S~2.
The first kind of plateau phase is unavoidable and should be observable
if a supernova with jets of the kind we have posited
is observed early enough.
The second kind can only exist if $b>>2$, and is probably
physically unrealizable for cases of significant polarization
as we discuss below.

    A representative set of the Table~\ref{table-1} predictions for $b=\infty$ 
with $\tau_{\rm rad}>>1$ are illustrated by the $b=\infty$ cases
in Figure~1 at times $t_{\rm red}$ significantly
less than 1.
As mentioned in \S~4.1, the $b=\infty$ cases can actually rise slightly,
but distinctly, above
the Table~\ref{table-1} predictions for $\Delta\theta\gtrsim30^{\circ}$
when $\tau_{\rm rad}$ declines toward the neighborhood of 1:
the actual localized maxima in the figure occur 
at $t_{\rm red}\approx 0.54$ ($\tau_{\rm rad}\approx 3.5$) for $\Delta\theta=30^{\circ}$
and 
at $t_{\rm red}\approx 0.66$ ($\tau_{\rm rad}\approx 2.3$) for $\Delta\theta=46^{\circ}$.
Declining $\tau_{\rm rad}$ tends to lower polarization by reducing
the number of photons scattered in the jet.
But, as discussed in \S~4.1, 
the competing effect of declining $\tau_{\rm off}$ (which increases
$w$ and thus tends to increase polarization) briefly dominates and 
the ideal radial phase high polarizing effect is approached.
The calculations show that this situation  
causes distinct polarization maxima to appear for 
sufficiently large $\Delta\theta$.
The $b=0$ cases are at all times lower in polarization than the $b=\infty$
cases of the same $\Delta\theta$ because 
the $b=0$ case $w$ factor is everywhere smaller, 
except all cases have zero polarization when
$t_{\rm red}=\infty$.
The $b=0$ cases decline monotonically with $t_{\rm red}$:  their only
time dependence comes through the $1-\exp(-\tau_{\rm rad})$ factor
which declines monotonically with time 
(see \S~3.5, eq.~[48]).

     Figure~2 shows the $b=1$ and $b=3$ cases for the same
$\Delta\theta$ values as used for Figure~1.
These cases further illustrate the effect of the $b$ parameter
on the time evolution.
As long as $\Delta\theta\tau_{\rm rad}/b>>1$ (i.e., at early
times with $\Delta\theta$ in radians, of course), the off-radial escape
of highly polarized photons is small and the polarization
is just as in the $b=0$ case:  i.e., $\tau_{\rm off}>>1$, and so
$w=q=1/4$.
This early phase is the first kind of polarization plateau phase
discussed at the end of \S~2.
 
   When $\Delta\theta\tau_{\rm rad}/b$ becomes small the ideal
radial phase effect become high and polarization begins to rise.
However, sometime thereafter $\tau_{\rm rad}$ becomes small and
the polarization falls as overall scattering declines.
Thus, we expect, as discussed at the end of \S~4.1, 
a polarization maximum in time evolution
when $\Delta\theta\tau_{\rm rad}/b\lesssim 1$ (i.e., there is high off-radial
escape probability for polarized photons), 
but $\tau_{\rm rad}\gtrsim 1$ (i.e., there
is still strong scattering in the jet). 
Figure~2 shows this is the case.
Note the polarization maxima are higher and shifted to earlier times
for the $b=3$ cases relative to the corresponding $b=1$ cases:
this is because for the $b=3$ cases there are
longer time periods starting earlier where the conditions
$\Delta\theta\tau_{\rm rad}/b\lesssim 1$ and $\tau_{\rm rad}\gtrsim 1$ hold
and a greater degree of the ideal radial phase effect is reached.
Note also that though polarization increases with greater $\Delta\theta$,
increased $\Delta\theta$ tends to shift the polarization maxima to
later times because again the condition $\Delta\theta\tau_{\rm rad}/b\lesssim 1$ has
to be approximately met for the maxima to occur.

   The maxima in Figure~2 do not reach the high polarizations and the second kind
of polarization plateaus obtained for corresponding $b=\infty$ cases 
(see Fig.~1) which shows that there
is never a time when $\Delta\theta\tau_{\rm rad}/b$ is negligible
in the evaluation of $\tau_{\rm off}$ while $\tau_{\rm rad}\gtrsim 1$.
For instance, the $b=3$, $\Delta\theta=30^{\circ}$ case reaches
a maximum polarization of $4.06\,$\% at $t_{\rm red}=0.74$
whereas the $b=\infty$, $\Delta\theta=30^{\circ}$ case had
polarization greater than $5\,$\% for $t_{\rm red}\lesssim 0.75$.

     The prediction of a period of rising polarization with time given by
the calculations for Figure~2 is qualitatively in agreement with the 
observations
of SN~1987A \citep{jeffery1991b} and SN~1999em \citep{leonard2001}.
The prediction of eventual decline of intrinsic polarization to a very low
level is believed to be in agreement with observations for SN~1987A
\citep{mendez1990, jeffery1991b}.
An eventual decline in supernova polarization with time must
occur in any model depending on electron scattering polarization
since the electron optical depth must decline eventually as
the supernova ejecta expands. 
Our polarization expression (eq.~[48] in \S~3.5) predicts that in
all cases polarization will start to fall when
$\tau_{\rm rad}$ declines to of order 1.

     Recall $\sim 4\,$\% polarization is a possible record for continuum
polarization for a core-collapse supernova (see \S~1.1).
The $b=3$, $\Delta\theta=30^{\circ}$ jet polarization shown in Figure~2 
reaches level. 
We consider bipolar jets with these parameters a physically plausible
case in which the jets are relatively narrow. 
It is satisfying that a plausible bipolar jet case can reach
the possible record polarization.

     In Figure~3, we show the polarization time evolution curves for
$\Delta\theta=10^{\circ}$ and a range of $b$ values
from $0$ to infinity.
As we would expect from our discussion of the Figure~2
curves above, the polarization maxima in Figure~3 get
higher and broader as $b$ is increased since the time period of
$\Delta\theta\tau_{\rm rad}/b$ small while $\tau_{\rm rad}\gtrsim 1$
is longer and there is a stronger ideal radial phase effect.
For $b\gtrsim 100$, there is the second kind of polarization plateau.

    Physically $b$ values over 50 are probably impossible since
they give $\Delta\beta_{\rm rad}=\beta_{\rm in}b>1$ 
(i.e., outer end jet speeds greater than light) even for
modest $\beta_{\rm in}=0.02$ (i.e., jet inner end speed
of about $6000\,{\rm km/s}$) which is of order of the
velocity of supernova photospheres for much of the
observable epoch.
(Recall as $\Delta\beta_{\rm rad}$ becomes more relativistic
our treatment becomes less adequate because we have neglected
relativistic effects in our formalism and because in the
present calculations we are neglecting the time-delay effect.)
Reducing $\Delta\theta$ allows the second kind of polarization plateau
to be reached with smaller $b$ values.
For example, $\Delta\theta=1^{\circ}$ in calculations otherwise
like those of Figure~3 causes the polarization plateau
to appear for $b\gtrsim 10$, but, as one would expect from
equation~(2) (\S~2) and equation~(48) (\S~3.5),
the polarization scales down everywhere by a factor of about 100
from the Figure~3 calculations:  the
$\Delta\theta=1^{\circ}$ curves 
have polarizations of order $0.01\,$\% or less which are currently
observationally insignificant.
(We believe that typically the lower limit of 
observationally significant polarization is perhaps $0.1\,$\%.
Depending on circumstances the lower limit could vary up and down from
this, of course.) 

     The upshot of the discussion in last paragraph is 
that the second kind of polarization plateau which is evident 
in Figures~1 and~3 is probably not physically realizable for
high polarization cases.
It seems unlikely that the inclusion of full relativistic effects
and time-delay effects will ease the requirement for very
high $b$ values in order to get the second kind of
polarization plateau for cases of significant continuum polarization.
Thus the second kind of polarization plateau is only a mathematical
prediction of the bipolar jet model for cases of
significant continuum polarization.
We do not expect it to be observed. 
Consequently, the very large and infinite $b$ value cases 
seen in Figure~3 and the infinite $b$ value cases of 
Figure~1 only illustrate mathematical trends of 
the bipolar jet model formalism.

     The other extreme from large $b$ is when $b$ is decreased
toward zero (see Fig.~3 again).
In this situation, the localized polarization maximum that
can occur near $t_{\rm red}=1$ (i.e., $\tau_{\rm rad}=1$)
can first cease to be a global maximum (see the $b=0.1$ curve) and then 
can disappear altogether (see the $b=0.01$ curve) as $b$ is reduced. 

    Figure~3 also illustrates that all curves approach the 
same functional behavior at late times where they decline like
$\tau_{\rm rad}=t_{\rm red}^{-2}$ (see eq.~[10] in \S~2) since 
$1-\exp(-\tau_{\rm rad})\to\tau_{\rm rad}$ at late times.
The late-time asymptotic curve is the
same for all cases, except for the $b=0$ curve.
This is because for $b>0$, the $\tau_{\rm off}$ quantity
goes to zero as time increases, and thus $w$ goes to 1 as time increases
(see eq.~[13] and~[15] in \S~2 and eq.~[42] and~[48] in \S~3.5).
The $b=0$ curve does not approach the same asymptotic curve
because $\tau_{\rm off}$ is always infinite, and thus $w=q=1/4$
for all times.

    The continuum polarization calculations we have done here
have been to demonstrate the range of polarization 
behaviors that might be expected from supernovae with bipolar jets
and to illustrate the mathematical properties of the polarization
formula (eq.~[48] in \S~3.5).
If one had observed continuum polarization as function of time 
(i.e., real time, not reduced time),
a fit to the observations could be done to determine values for
$\theta$, $\Delta\theta$, $b$, and 
the $\tau_{\rm rad}t_{\rm 0}^{2}$ parameter (see \S~2, eq.~[9]).
Unfortunately, the degeneracy of the parameters which we discussed
at the end of \S~3.5 would make it difficult to
determine a unique set of parameters.

\subsection{Polarization Line and Continuum Features}

     Polarization spectra would have exactly constant polarization
without the redshift effect of the jets (see \S~3.3, esp.~eq.~[30]). 
With the redshift effect, P~Cygni line flux features tend to give rise
to inverted P~Cygni polarization profiles:  high polarization
at the blueshifted trough of P~Cygni flux profile
and low polarization at the 
near-line center maximum of a P~Cygni flux profile.
This kind of line polarization profile is also produced by other
supernova polarization models (see \S~1.1 for other supernova
polarization models) and is,
in fact, present or partially present 
for some observed supernova P~Cygni lines (e.g.,
\citealt{jeffery1991b}, \citealt{wang2001}, and \citealt{leonard2001}).
That the bipolar jet model can yield such a profile is a success.
However, the observed profiles often show shifts in
position angle which the strictly axisymmetric
bipolar jet model cannot yield.
Thus, the axisymmetric bipolar jet asymmetry even if it exists
cannot be the only asymmetry in many cases and maybe in all cases. 

     The inverted P~Cygni profiles in the bipolar jet model arise 
as follows.
Because of the redshift effect 
the jet-scattered, polarized flux originates 
in direct flux spectrum from a bluer part of that spectrum than
the part of that spectrum to which it (i.e., the jet-scattered, polarized flux)
is added to make up the total flux spectrum. 
Now if this bluer flux is redshifted from
a continuum region to another continuum region there will be a
certain scattered flux to direct flux ratio and a certain polarization
in the second region.
If the bluer flux is redshifted from a continuum region to
P~Cygni trough region, then there will be a higher
scattered flux to direct flux ratio and higher polarization in
the second region:  the higher 
polarization is because there is less dilution of the scattered, polarized
flux by unpolarized direct flux.
If the bluer flux is redshifted from a continuum region to 
P~Cygni emission peak region, then there will be a lower
scattered flux to direct flux ratio and lower polarization in the second
region: the lower 
polarization is because there is more dilution of the scattered, polarized
flux by unpolarized direct flux.
The net result of scattered continuum flux redshifted to a P~Cygni
line, is an inverted P~Cygni polarization profile associated with
P~Cygni flux profile:  higher
than continuum polarization in the trough region;  lower than
continuum polarization in the emission region. 

    Now P~Cygni line widths are caused by Doppler shifts on
scattering in the fast moving ejecta at and beyond the photosphere: 
the Doppler shifts correspond typically to velocities of order 1 or 2
times the photospheric velocity $\beta_{\rm ph}$:  
the shifts can be larger for very strong lines.
Since we assume the jets and their main scattering
region at $\beta_{\rm scat}$ to be well outside the photosphere
(i.e., $\beta_{\rm scat}>>\beta_{\rm ph}$), the
jet redshift is usually going to be larger than the typical P~Cygni line
width. 
Thus, jet-scattered flux redshifted into a P~Cygni line wavelength
region will usually come from an unrelated bluer part of the spectrum 
which could be continuum or could be complicated by other P~Cygni lines 
possibly overlapping each other.
But if the P~Cygni line is a strong one, then usually
the redshifted, jet-scattered flux that is added to it
will be more continuum-like than the P~Cygni line itself.
Thus for strong P~Cygni lines or P~Cygni lines sufficiently
well isolated by continuum regions, one usually expects
inverted P~Cygni polarization profiles by the mechanism discussed in
the last paragraph.
Admittedly, a complex direct flux spectrum with
many overlapping P~Cygni will give rise to a complex
polarization spectrum in which the inverted P~Cygni profiles
might not be clearly present.

     Purely continuum polarization features can arise if the 
jet is very fast:  i.e.,
$\beta_{\rm scat}>>\beta_{\rm ph}$.
This is because there will be very large redshifts
and variations in the direct flux continuum level will affect the
polarization spectrum.
For example, say that the jet redshift moves direct flux from the 
ultraviolet (UV)
region to the blue part of the optical and from the
blue part to the red part of the optical.
Now the UV flux of a supernova is very low compared 
to the optical in the observable epoch after very early
times because of strong line-blanketing.
In this case, the jet-scattered UV flux that is redshifted 
into the blue part of the optical will be comparatively
very low and even if highly polarized might
be insufficient to cause significant polarization of the net flux. 
On the other hand, in the red part of the optical, the polarization
should be higher than in the blue part 
since the jet-scattered flux from the blue part is shifted there. 
Thus, polarization should rise strongly across the optical in this
example situation.

\subsection{Polarization Spectrum Calculations}

    Using the expressions for relative flux $R$ and polarization $P$
(eq.~[47] and~[48]
in \S~3.5) and a set of the 5 independent basic parameters 
for the bipolar jet model, one can calculate 
the direct-plus-jet spectrum 
and polarization spectrum for any synthetic spectrum calculated
in spherical symmetry.
One can also take any observed flux spectrum and a set of
the 5 independent basic parameters and calculate a direct spectrum
(see \S~3.5 for the method) and a polarization spectrum.
If one has an observed polarization spectrum, one can do
a fit to that spectrum by adjusting the parameters.
Unfortunately, the effect of parameters is very degenerate since
they combine in a complex way to control the level of polarization as
discussed at the end of \S~3.5.
An analysis of the time evolution of the continuum polarization might 
allow one to at least partially break the degeneracy and
obtain values for $\theta$,
$\Delta\theta$, $b$, and $\tau_{\rm rad,0}t_{0}^{2}$
(see \S~3.5 and~4.2).
An analysis of the polarization spectrum would then allow a
determination of $\beta_{\rm in}$ parameter since it has a strong
effect on the redshift effect (see eq.~[14] in \S~2
and eq.~[30] in \S~3.3).

     As a demonstration we have calculated the direct
spectrum and polarization spectrum for an observed spectrum
of SN~Ib~1999dn (in NGC~7714) from
17~days after $V$ maximum (i.e., about 37~days after explosion
assuming a rise time of about 20~days:  see \S~1.2).
The observed spectrum is from \citet{matheson2001}.
The photospheric velocity $\beta_{\rm ph}$ of the supernova
at that epoch was $0.02$ (i.e., $v_{\rm ph}=6000\,{\rm km/s}$) 
(BBK).  
For the 5 independent basic parameters we have chosen 
$\theta=90\,$\% (for simplicity and high polarization),
$\Delta\theta=20\,$\% (for high polarization and yet still
to have a relatively narrow jet:  it is also the fiducial value
from \S~2),
$b=1$ (for simplicity:  it is also our fiducial value from \S~2),
$\beta_{\rm in}=0.06$ (for simplicity since it is
exactly 3 times the photospheric velocity and also for a significant
redshift effect),
and
$\tau_{\rm rad}=0.67455$ (for relatively high polarization in
the radial phase of the jets and to
obtain the fiducial mass of \S~2:  see below).

   The dependent parameters implied by the independent basic parameters
are $t_{\rm red}=1.2176$, $\tau_{\rm mean}=0.300$,
$\Delta\beta_{\rm rad}=0.06$,
$\tau_{\rm off}=0.340$, and $\beta_{\rm scat}=0.0867$.
Note $\omega_{\mu}=0$ for $\tau_{\rm off}\leq 1$ by our
prescription equation~(44) (\S~3.5), but its value is
irrelevant since $\theta=90\,$\% (see eq.~[48] in \S~3.5).
If we neglect the redshift effect and set
$[f_{\rm dir}(\lambda_{+}')(1-w_{\mu}\mu)
 +f_{\rm dir}(\lambda_{-}')(1+w_{\mu}\mu)]
/f_{\rm dir}(\lambda)=2$,
$R$ and polarization 
are, respectively, 0.0238 and $1.56\,$\%:
these are characteristic $R$ and polarization values for
the calculation. 
Interpolating from Figure~2, we know for the specified
$\tau_{\rm rad}$ and reduced time that
the continuum polarization would be declining
from its time evolution maximum. 

    If we choose $\mu_{e}=4$ and $t=37\,$days, then equations~(7)
and~(8) (\S~2) imply that each jet has mass $0.1\,M_{\odot}$
and characteristic kinetic energy 
$0.724\times10^{51}\,{\rm ergs}=0.724\,$foe.
This mass and kinetic energy were the fiducial jet values of \S~2
and our choice of precisely $0.67455$ for $\tau_{\rm rad}$
was made to yield them.

     Figure~4 shows the observed flux spectrum and the calculated
direct flux spectrum.
The direct spectrum was calculated using a $R$ value which in
turn was calculated from the observed flux by the method
suggested in \S~3.5.
Since the characteristic $R$ value for the calculation is small
($0.0238$:  see above), this method is quite accurate.
The scale of the spectra is arbitrary.
The spectra have been corrected for the redshift
due to the host galaxy heliocentric velocity
($2798\,{\rm km/s}$) and
for the host galaxy Galactic foreground reddening ($E_{\rm B-V}$=0.052)
using the reddening law of \citet{cardelli1989} with the
revision of \citet{o'donnell1994}.
Both host galaxy correction parameters were obtained from the NED 
database ({\tt http://nedwww.ipac.caltech.edu/}).

     Since the direct spectrum is on average only about $2.4\,$\%
lower than the observed spectrum, the two spectra virtually
overlap.
The effect of the jet-scattered flux on the observed spectrum
is probably negligible compared to observational error.
Having a negligible effect on an observed spectrum
at the same time as producing significant polarization was,
of course, a main desideratum from the bipolar jet model since
this desideratum would resolve the potential paradox discussed in \S~1.2. 

    Figure~5 shows the calculated bipolar jet model
polarization spectrum overlaid on the observed flux spectrum.
Note that the polarization spectrum can only extend as far
to the blue as the jet-redshifted blue end of the flux spectrum.
Inverted P~Cygni polarization profiles associated with
flux spectrum P~Cygni lines are evident:  the one
associated with the H and K Ca II P~Cygni line (trough
centered at $\sim 3750\,$\AA) is the most clear example.
The line blending in the flux spectrum, of course, somewhat
distorts the inverted P~Cygni polarization profiles.
The largest polarization maxima, not counting the far red end of
the spectrum, reach over $4\,$\%.
The complex high polarization features at the red end of the spectrum
rise higher:   the highest is about $9\,$\%.  
We cut off these high polarization features in the figure to preserve
the clarity of the rest of the polarization spectrum.
In any case, the red end of the flux spectrum from which the 
these high polarization features are partially
determined is probably uncertain.

    For Figure~6, we have changed the independent basic parameter
$\beta_{\rm in}$ to $0.12$.   
The changed dependent parameters are $\Delta\beta_{\rm rad}=0.12$
and $\beta_{\rm scat}=0.173$.
The implied mass of each jet is changed from the 
$0.1\,M_{\odot}$ to $0.4\,M_{\odot}$:
this follows from holding the ratio $M/\beta_{\rm in}^{2}$
fixed in equation~(7) (see \S~2) so that the other values
in that equation can stay unchanged.
This mass is a relatively large fraction of 
a typical SN~Ib where total ejecta mass is perhaps
only $4\,M_{\odot}$ or less (e.g., \citealt{hachisu1991}).  

    In the Figure~6 calculation, the wavelength 
of jet-scattered flux is increased by a factor of 
$1.173$ (see \S~3.3, eq.~[29] with $\mu=0$) which is so large
that the jet-scattered flux comes from direct flux spectrum
wavelength regions where the continuum levels
can be rather different that those of the wavelength regions 
to which the scattered flux is added to give the net spectrum.
A result of this large redshift is that polarization at the blue end
of the optical is somewhat reduced in overall continuum trend 
relative the Figure~5 case.
The polarized flux in the blue part originates in even
bluer flux of the direct spectrum:  but since the
direct spectrum declines blueward on average, there is less polarized
flux at a given wavelength than if the redshift were smaller.

    Figure~6 like Figure~5 shows inverted P~Cygni polarization
profiles.

\section{Conclusions and Final Discussion} \label{conclusions}

      We have proposed a bipolar jet model for supernova polarization.
The motivation for this model is to resolve the potential
paradox set by BBK analysis which suggests strong spherical
symmetry for Type~Ib supernovae (see \S~1.2)
and the expectation that Type~Ib
supernovae should be highly polarized in the continuum
up to perhaps $\sim 4\,$\% (see \S~1.1).
For the model, we have derived simple, non-relativistic
approximate analytic formulae for relative scattered flux $R$
(i.e., for the ratio of scattered flux 
to direct flux from the bulk supernova) and for polarization
(see eq.~[47] and~[48] of \S~3.5).
The polarization formula qualitatively reproduces two main
features observed for at least some supernovae:
(1)~the rise of continuum polarization to a maximum and then a
decline with time (see \S~4.2);  
(2)~the inverted P~Cygni polarization profiles of lines
(see \S~4.4).
The level of continuum and line polarization obtained reaches 
that observed for supernovae (i.e., of order a percent or
a few percent:  see \S~1.1) 
for plausible jet model parameters:  in fact, one set of plausible parameters
does allow continuum polarization to reach $\sim 4\,$\% (see \S~4.2). 
The position of angle polarization is a constant for the bipolar jet model
since it is a strictly axisymmetric model and
the model does not allow $90^{\circ}$ shifts in position angle which
are possible for axisymmetric models in general.
Since position angle variation is observed, the model cannot
account for all the polarization of all supernovae. 

    The bipolar jet model predicts that two kinds of
the polarization plateau phases in time can exist (see \S~2).
The first kind of plateau phase is unavoidable and should be observable
if a supernova with jets (of the kind we have posited)
is observed early enough:  we show
cases of this plateau phase in Figures~1, 2, and~3 in \S~4.2.
The second kind can only exist if $b>>2$ (see \S~2):  we show cases
with and without the second kind of plateau phase 
in Figures~1, 2 and~3 in \S~4.2.
The second kind is probably physically unrealizable for jets
wide enough to give observationally significant continuum polarization
(see \S~4.2). 
We have not yet examined existing supernova polarization data to
find evidence for the kinds of polarization plateaus predicted by
the bipolar jet model.

      We derived the analytic formulae for relative scattered flux
and polarization using very simplifying assumptions.
We have treated the electron optical depth structure of the jets very
crudely and neglected any line or other opacity in them.
The geometry of the supernova-jet structure has also been treated
crudely (see \S~2).
The time-delay effect (see \S~3.2)
can be included in the model, but it can be
neglected for simplicity as we have done in deriving the
analytic formulae and in the demonstration
calculations.
We have assumed the bulk supernova (i.e., the supernova
not counting the jets) is 
spherically symmetric: this cannot be completely true although to resolve the
potential paradox it must be roughly true.

    There are 5 independent basic parameters for the bipolar jet model:
$\theta$ (or $\mu=\cos\theta)$, 
$\Delta\theta$, $\tau_{\rm rad}$, $\beta_{\rm in}$, and $b$
(see \S~3.5) which unfortunately are rather degenerate particularly
for the analysis of a single polarization epoch.
For an analysis of the time evolution of the polarization 
the $\tau_{\rm rad}t_{\rm 0}^{2}$ parameter (see \S~2, esp. eq.~[9])
can replace $\tau_{\rm rad}$ as one 
of the 5 independent basic parameters.
In an analysis in which the time-delay effect is included, $t_{\rm red}$
(see \S~3.2) replaces $\tau_{\rm rad}$ as one of the
5 independent basic parameters.

     We have neglected relativistic effects in our formalism.
If $\beta_{\rm scat}$ becomes sufficiently large
(e.g., $\gtrsim 0.1$), then this neglect could be a significant limitation
of the formalism since relativistic effects will be important.
These effects could significantly reduce the
calculated supernova polarization relative
cases that neglect them, but have otherwise the same parameters.
The essential reason is that energy is lost from scattered flux by
Doppler shifting into
the jet frame that is not regained Doppler shifting back to the
rest frame of the supernova center.
If the jet-scattered flux decreases because of this process, then
the amount of polarized flux decreases and net polarization
decreases.
Thus, our formulation may overestimate polarization for jets
that are too relativistic. 
In future work we intend to investigate relativistic effects 
in the bipolar jet model.

     Because of our simplifying assumptions,
the derived formulae can only be expected to
semi-quantitatively reproduce the effects of bipolar jets and,
of course, the parameters used in the formulae will have only a crude
relation to the parameters of actual bipolar jets which we expect
to have rather complex structure.
The formulae should be tested by Monte Carlo calculations.
Such tests would probably show how to adjust the formulae to
improve their quantitative accuracy.
In any case, variations on the formulae we 
have presented are clearly possible.
In the analysis of actual supernova polarimetric data, some
variations may turn out to be obviously more appropriate than 
formulae we have given.

      The advantage of the analytic formulae is that they allow 
a straightforward, intelligible analysis of polarimetry data by
fitting the data and they allow
a simple first-order approach to testing the adequacy 
of the bipolar jet model.
We have verified the adequacy of the model to some
degree in this paper:  see the first paragraph of this section.
If the model's adequacy is further verified, then
more a more elaborate testing procedure for the model should 
be considered.
For example the calculation of polarization spectra using
Monte Carlo radiative transfer for a bipolar jet model whose
structure is produced---if it is feasible---by a realistic 
hydrodynamic calculation.

     In this paper, we have only considered the bipolar jet model
with SNe~Ib in mind.
The bipolar jet model may apply generally to core-collapse supernovae.
On the other hand it may only to apply to some:  i.e., some
SNe~Ib and, perhaps, some of their smaller-envelope kin, Type~Ic supernovae.
In future work we hope to test the bipolar jet model using 
polarimetry and spectropolarimetry from all types of core-collapse
supernovae.

      As final word, one must emphasize that neither the
bipolar jet model nor the other models of supernova asymmetry
(see \S~1.1) are exclusive.
Several kinds of asymmetry can be present for a single supernova.
For example, Type~Ic supernova SN~2002ap shows a component of polarization 
that can modeled by a jet \citep{kawabata2002, leonard2002}.
There is certainly another polarization component present though
in SN~2002ap perhaps due to a inner ejecta asymmetry of order
$10\,$\% in axis ratio projected on the sky 
\citep{kawabata2002, wang2003}.
The late direct HST imagining of SN~1987A confirms that in one supernova
at least, there is a large inner ejecta asymmetry:  the 
major to minor axis radio of images is
of order $1.5/1$ \citep{wang2002}.
It may be that such inner ejecta asymmetry (caused perhaps by
non-relativistic, never-emergent bipolar jets:
\citealt{hoeflich2001}) is the main kind of SN~1987A asymmetry.
However, a re-analysis of early SN~1987A speckle
interferometry \citep{nisenson1999} suggests that relativistic
jets (with perhaps $\beta\sim 0.8$) did emerge from SN~1987A in addition 
to other asymmetries.
These suggested jets, however, may have been too relativistic to
have contributed significantly to SN~1987A's polarization:
this a subject for further study.



\acknowledgments

I thank David Branch and Eddie Baron for stimulating this work
and their comments on it.
Support for the paper was provided by the Departments of Physics
of New Mexico Tech and the University of Nevada, Las Vegas,
and the Department of Physics and Astronomy
of the University of Oklahoma.
This research has made use of the NASA/IPAC Extragalactic Database 
(NED) which is operated by the Jet Propulsion Laboratory, 
California Institute of Technology, 
under contract with the National Aeronautics and Space Administration.

\clearpage


\begin{figure}
\plotone{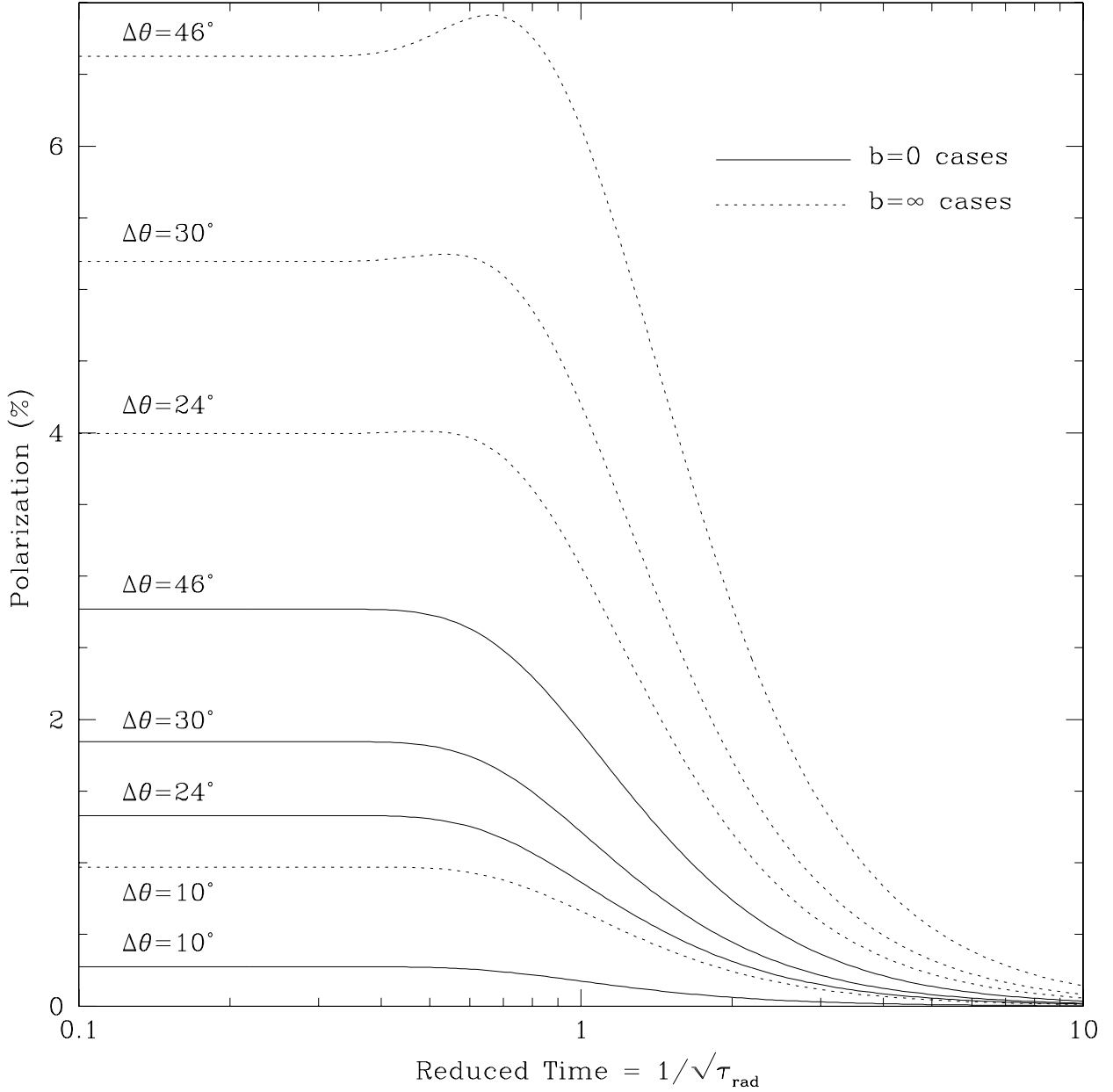}
\caption{
Continuum polarization as a function of reduced time
for a range of $\Delta\theta$ values as specified
on the figure, and $b=0$ and $b=\infty$.
We have set
$\theta=90^{\circ}$
and the direct supernova flux is wavelength independent.
The later condition means that
$[f_{\rm dir}(\lambda_{+}')(1-\omega_{\mu}\mu)
 +f_{\rm dir}(\lambda_{-}')(1+\omega_{\mu}\mu)]
/f_{\rm dir}(\lambda)]=2$, and
the $\beta_{\rm in}$ parameter is not used
and is left unspecified.
\label{fig1}}
\end{figure}

\clearpage 

\begin{figure}
\plotone{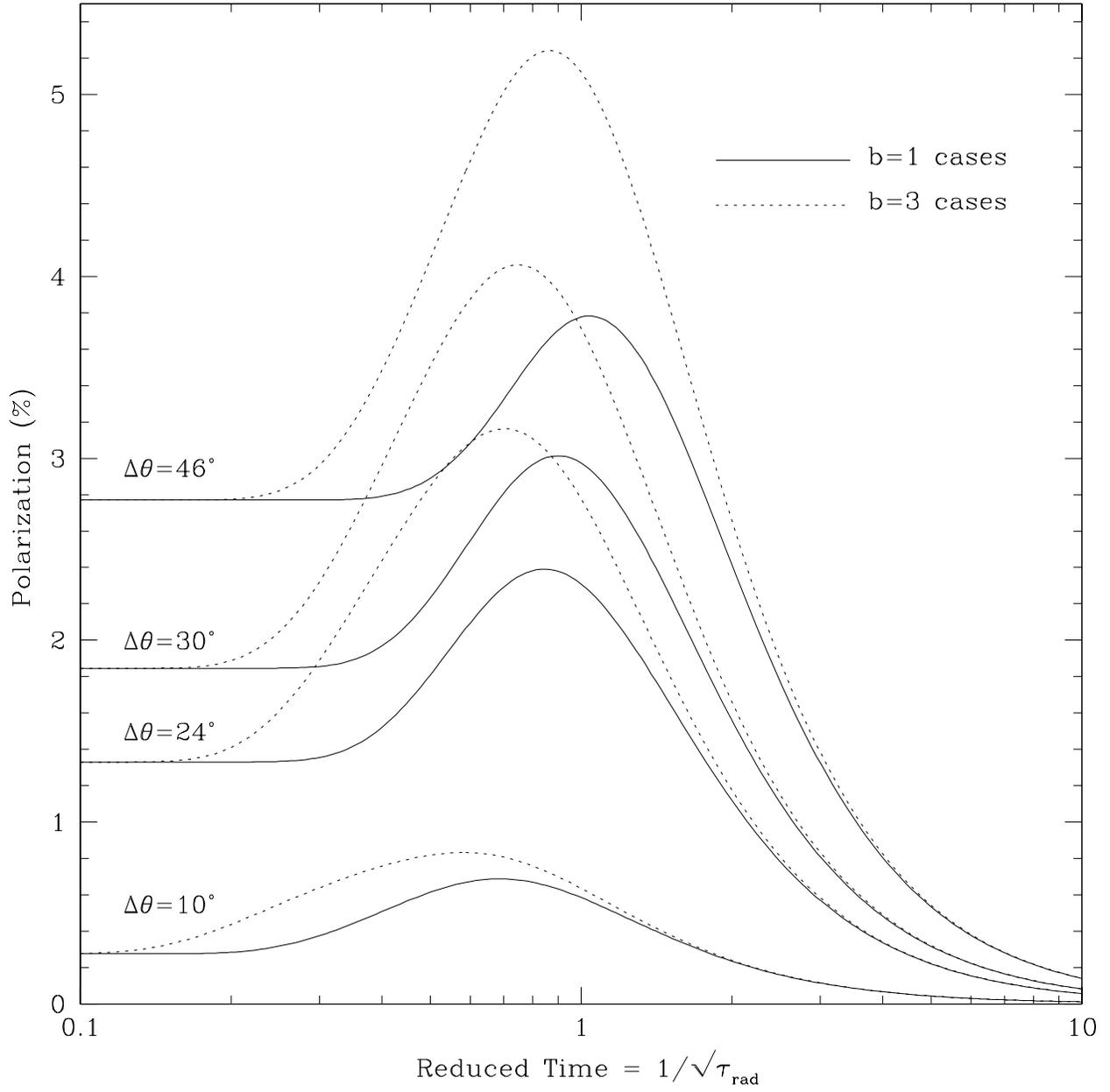}
\caption{
The same as Figure~1, except the polarization curves 
are for $b=1$ and $b=3$.
\label{fig2}}
\end{figure}

\clearpage

\clearpage

\begin{figure}
\plotone{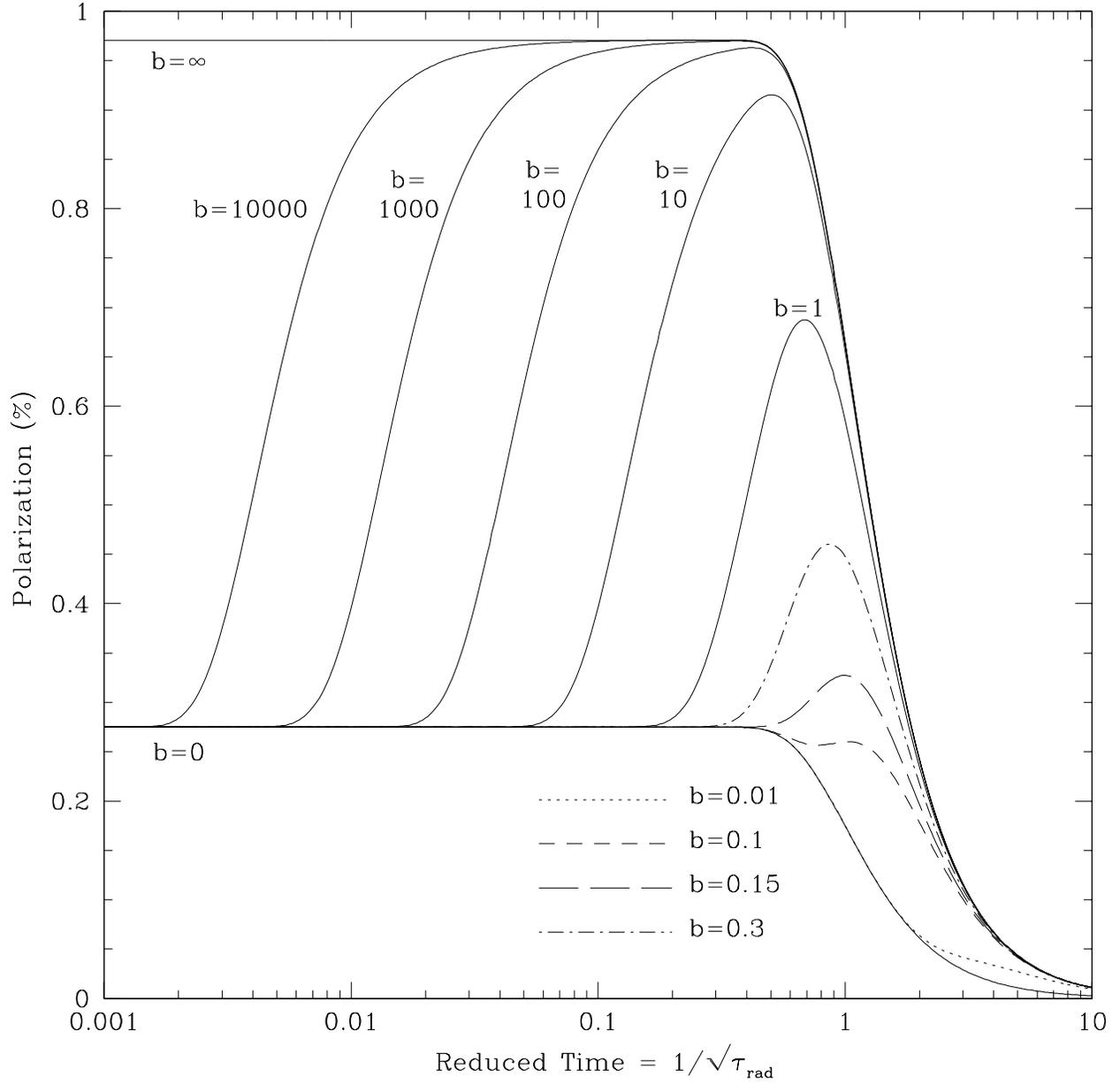}
\caption{
The same as Figure~1, except that 
$\Delta\theta$ is set to $10^{\circ}$ and
a range of $b$ values as specified on the figure
are used.
\label{fig3}}
\end{figure}

\clearpage

\begin{figure}
\plotone{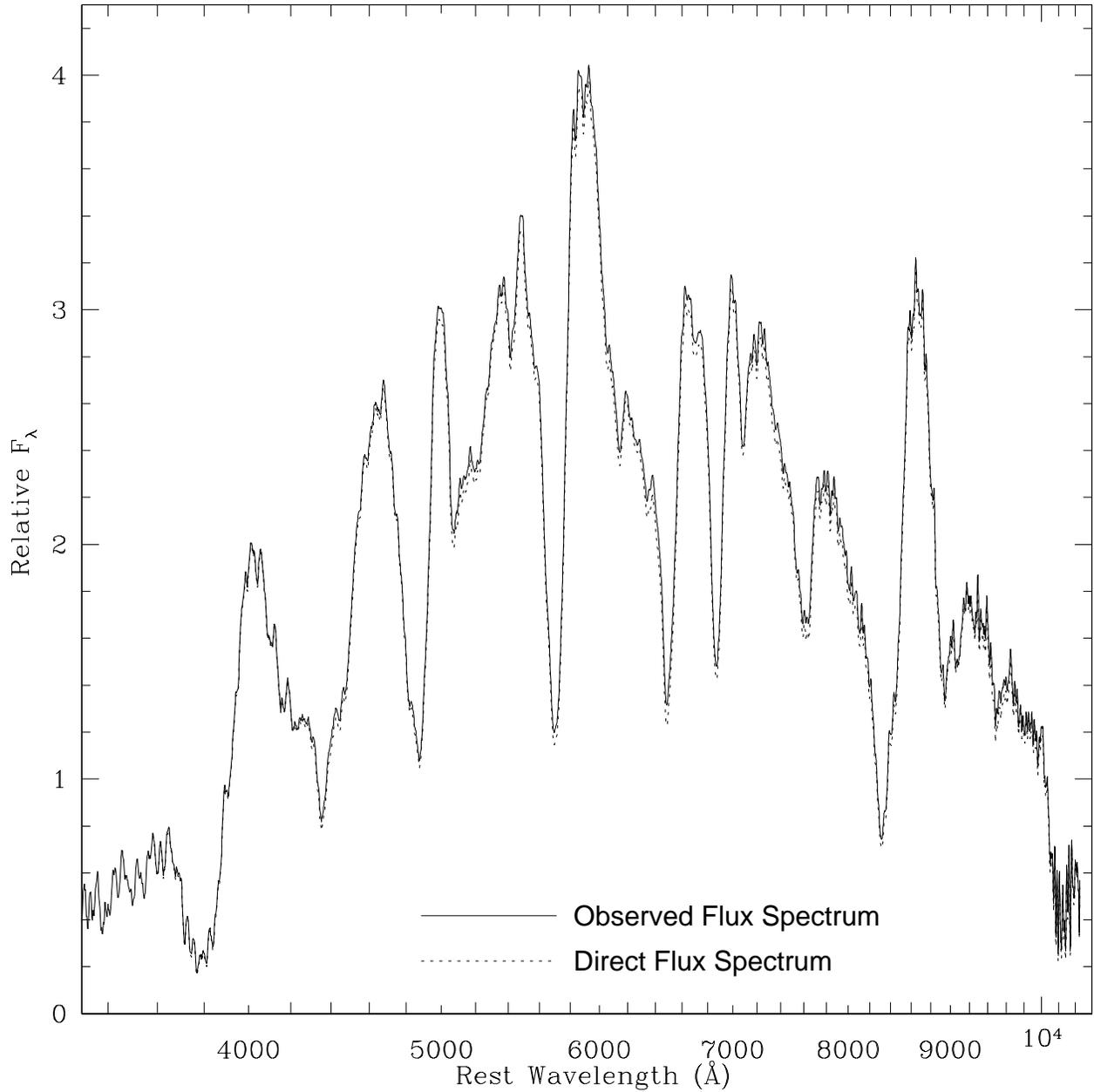}
\caption{
The observed flux spectrum and bipolar jet model direct flux spectrum
of SN~1999dn for day 17 past $V$ maximum.
The scale of the spectra is arbitrary. 
The bipolar jet model parameters are specified in the
text (see \S~4.4).
Both spectra have been corrected for heliocentric
velocity redshift and foreground Galactic extinction.
\label{fig4}}
\end{figure}

\clearpage

\begin{figure}
\plotone{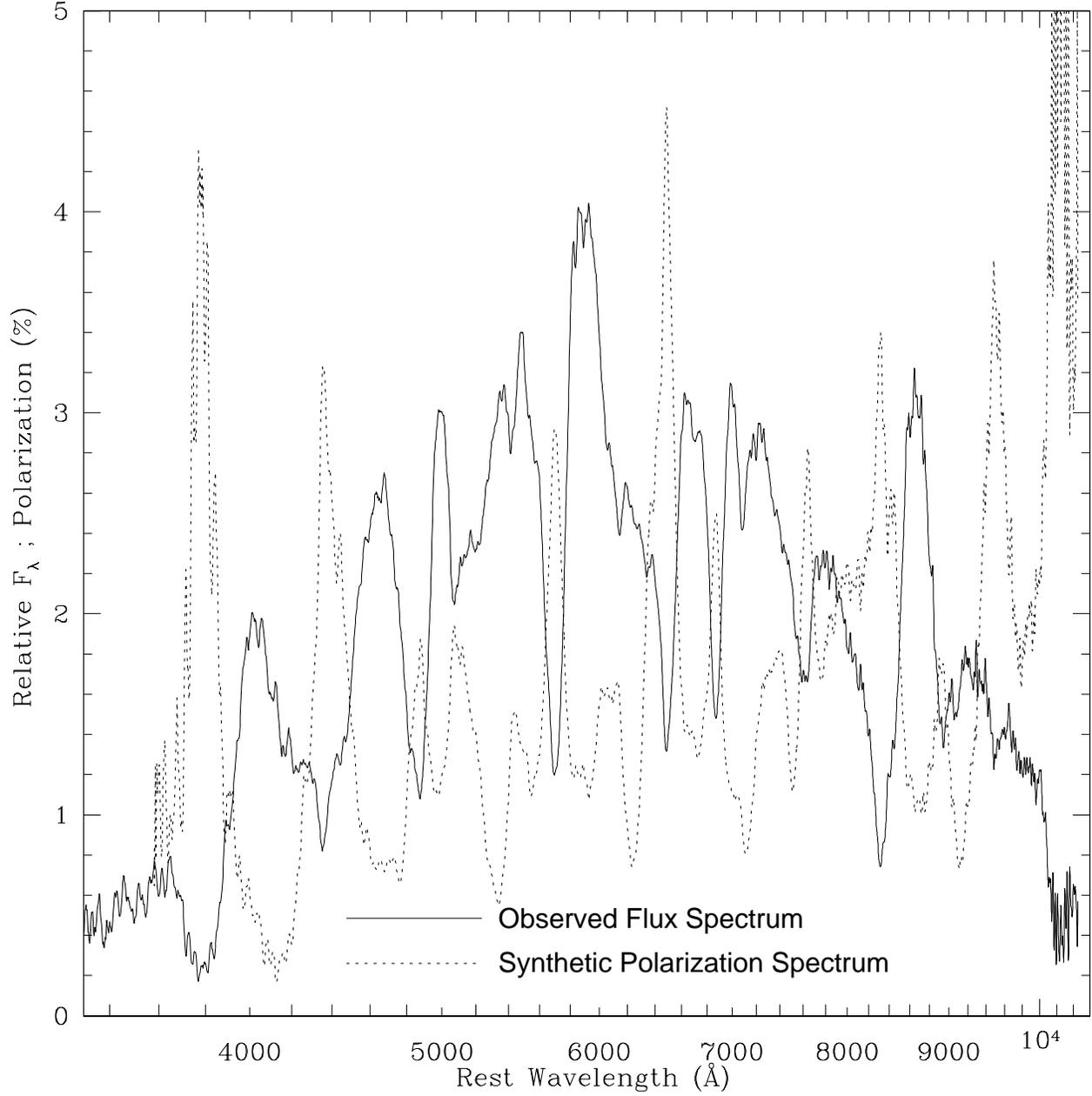}
\caption{
The bipolar jet model polarization spectrum for the observed
flux spectrum of Fig.~4 which is shown overlaid.
The bipolar jet model parameters are the same as for Fig.~4
and are specified in the text (see \S~4.4).
\label{fig5}}
\end{figure}

\clearpage

\begin{figure}
\plotone{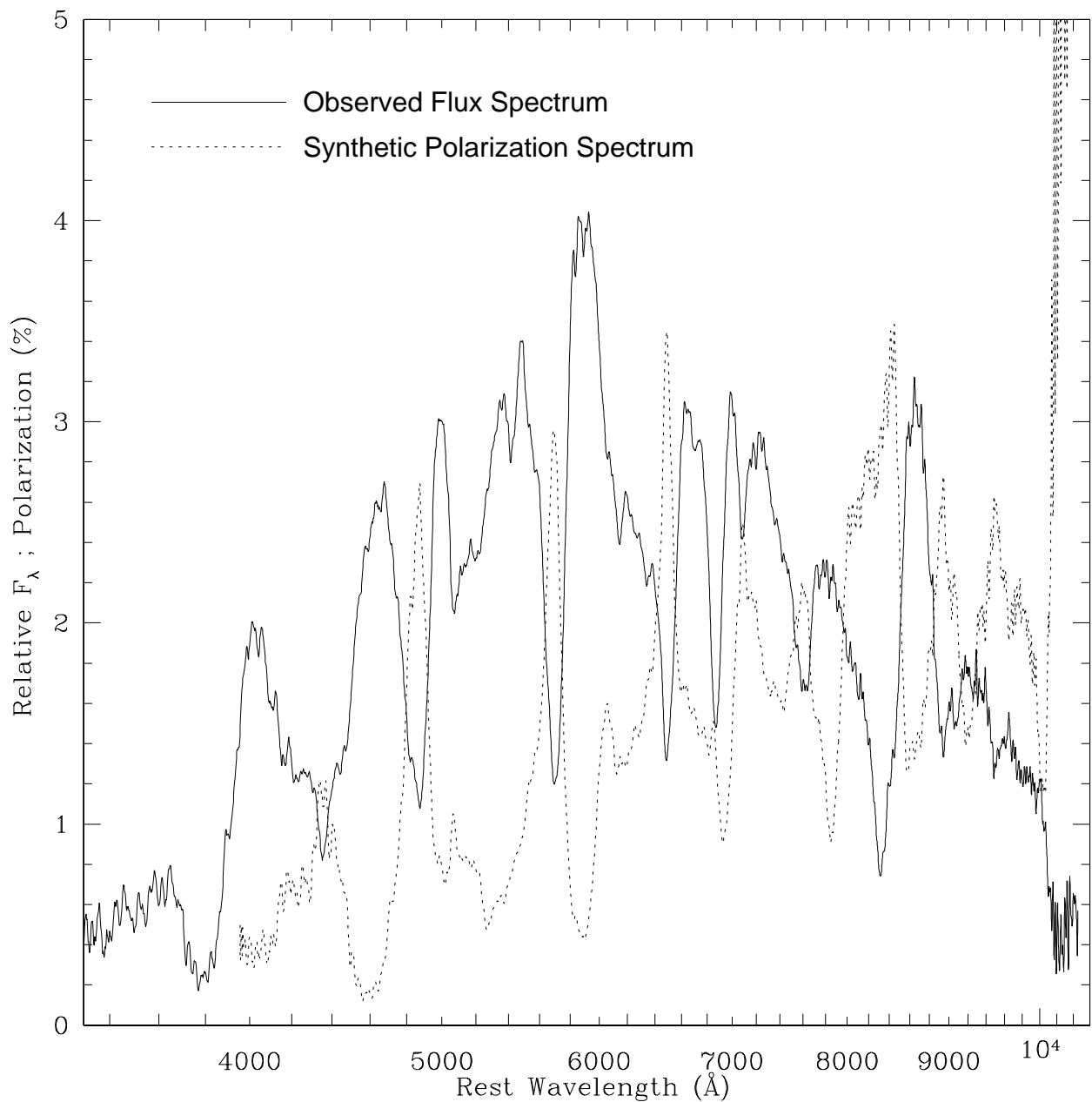}
\caption{
The bipolar jet model polarization spectrum for the observed 
flux spectrum of Fig.~4 which is shown overlaid. 
The bipolar jet model parameters are as for Figs.~4 and~5 (see text
\S~4.4), 
except the $\beta_{\rm in}$ parameter has been doubled from
$0.06$ to $0.12$.
This greatly increases the redshift effect of the jets.
\label{fig6}}
\end{figure}

\clearpage






\clearpage

\begin{deluxetable}{crrrrrrrrrrr}
\tabletypesize{\scriptsize}
\tablecaption{Runs of
High Continuum Polarization
and Ancillary Quantities as Functions $\Delta\theta$
for the Radial Phase of the Bipolar Jet Model
\label{table-1}}
\tablewidth{0pt}   
\tablehead{
 \colhead{$\Delta\theta$}
&\colhead{$\Omega_{\rm fr}(\Delta\theta)$}
&\colhead{$\xi(\Delta\theta)$}
&\colhead{$w$}
&\colhead{$R$}
&\colhead{P}  \\
\colhead{($^{\circ})$}   &&&&& \colhead{(\%)}     \\ 
}
\startdata
 \colhead{\hfill 0} &\colhead{0} &\colhead{1} &\colhead{1} &\colhead{0} 
    &\colhead{0}  \\
 \colhead{\hfill 1} & $7.6\times10^{-5}$ & 0.9998 & 0.9870 & $1.1\times10^{-4}$ & 0.0113 \\
 \colhead{\hfill 3} & $6.9\times10^{-4}$ & 0.9982 & 0.9617 & 0.0010 & 0.0986 \\
 \colhead{\hfill 7} & 0.0037 & 0.9901 & 0.9137 & 0.0058 & 0.5029 \\
 10 & 0.0076 & 0.9798 & 0.8799 & 0.0119 & 0.9708 \\
 15 & 0.0170 & 0.9549 & 0.8272 & 0.0270 & 1.9657 \\
 20 & 0.0302 & 0.9207 & 0.7790 & 0.0486 & 3.0939 \\
 24 & 0.0432 & 0.8871 & 0.7433 & 0.0704 & 3.9944 \\
 30 & 0.0670 & 0.8270 & 0.6943 & 0.1107 & 5.1942 \\
 40 & 0.1170 & 0.7053 & 0.6231 & 0.1975 & 6.4399 \\
 46 & 0.1527 & 0.6224 & 0.5860 & 0.2606 & 6.6262\tablenotemark{a} \\
 50 & 0.1786 & 0.5643 & 0.5634 & 0.3069 & 6.5166 \\
 60 & 0.2500 & 0.4135 & 0.5132 & 0.4359 & 5.5421 \\
\enddata


\tablenotetext{a}{
Maximum continuum polarization (for the chosen parameters)
as a function of $\Delta\theta$.
To better numerical accuracy it is $6.6271108\,$\% 
at $45.600748^{\circ}$.
}

\tablecomments{
The predictions were calculated using equation~(47)
for $R$ and equation~(48) for $P$ assuming a wavelength
independent direct flux from the bulk supernova.
The fixed parameter values for the predictions are
$\theta=90^{\circ}$, $b=\infty$,
and $\tau_{\rm rad}>>1$ (see text, \S~4.1).
The $\beta_{\rm in}$ parameter is not specified
since it is irrelevant to the calculation.
}

\end{deluxetable}


\clearpage




\end{document}